\documentclass[a4paper,10pt,twocolumn,accepted=2025-02-26]{quantumarticle}
\pdfoutput=1
\usepackage{graphicx} %
\usepackage[numbers]{natbib}

\usepackage[utf8]{inputenc}
\usepackage[english]{babel}
\usepackage[T1]{fontenc}
\usepackage{amsmath, amssymb}
\usepackage{braket}
\usepackage{booktabs}

\usepackage[colorlinks=true,     linkcolor=blue, citecolor=blue, filecolor=blue, urlcolor=blue]{hyperref}

\newcommand{\add}[1]{#1}
\newcommand{\del}[1]{}
\newcommand{\rep}[2]{#2}

\title{Chaos and magic in the dissipative quantum kicked top}

\author{G.~Passarelli}
\orcid{0000-0002-3292-0034}
\email{gianluca.passarelli@unina.it}
\thanks{corresponding author}
\affiliation{Dipartimento di Fisica ``E.~Pancini'', Universit\`a di Napoli Federico II, I-80126 Napoli, Italy}
\author{P.~Lucignano}
\email{procolo.lucignano@unina.it}
\affiliation{Dipartimento di Fisica ``E.~Pancini'', Universit\`a di Napoli Federico II, I-80126
Napoli, Italy}
\author{D.~Rossini}
\email{davide.rossini@unipi.it}
\affiliation{Dipartimento di Fisica dell'Universit\`a di Pisa \& INFN Sezione di Pisa,
  Largo B. Pontecorvo 3, Pisa, I-56127, Italy}
\author{A.~Russomanno}
\email{angelo.russomanno@unina.it}
\affiliation{Dipartimento di Fisica ``E.~Pancini'', Universit\`a di Napoli Federico II, I-80126
Napoli, Italy}

\begin{document}

\begin{abstract}
    We consider an infinite-range interacting quantum spin-1/2 model, undergoing periodic kicking and dissipatively coupled with an environment. In the thermodynamic limit, it is described by classical mean-field equations that can show regular and chaotic regimes. At finite size, we describe the system dynamics using stochastic quantum trajectories.
    We find that the asymptotic nonstabilizerness (alias the {\bf{\emph{magic}}}, a measure of quantum complexity), averaged over trajectories, mirrors \add{to some extent} the classical chaotic behavior, while the  entanglement entropy  has no relation with chaos in the thermodynamic limit.
\end{abstract}

\maketitle

\section{Introduction}

A deterministic dynamics may give rise to chaos: an aperiodic dynamics that strongly depends on the initial conditions~\cite{Gleick:1987,Schuster,Ott,baba:book}. In Hamiltonian many-body systems, this phenomenon is crucial to develop thermalization~\cite{Lichtenberg,Berry_regirr78:proceeding} and has been extended to the quantum domain~\cite{d2016quantum,Berry_Les_Houches}. Pretty much different from the Hamiltonian chaos is the dissipative one, where dynamical structures as strange attractors~\cite{Ott} appear, as well as universal routes to chaos, like the period-doubling cascade~\cite{Feigenbaum}.
Quantum dissipative systems with a well defined classical limit governed by a chaotic dynamics have been also studied~\cite{Dittrich_1988, braun, Benenti_2002, PhysRevLett.94.164101, PhysRevLett.95.164101, Carlo_2017, Bergamasco_2023, ferrari2023steadystate}. 

In this manuscript, we consider an infinite-range interacting quantum spin-1/2 system undergoing periodic driving and coupled to an external environment, which we dub the {\it dissipative kicked top}\footnote{The emergence of chaos in the mean-field limit of a strictly similar classical model, bearing the same name, has been addressed in Ref.~\cite{PEPLOWSKI1988408}. The corresponding finite-size quantum model has been studied using quantum-state-diffusion trajectories~\cite{N_Gisin_1992, J_Iwaniszewski_1995}, with a focus on expectations averaged over randomness, which is very different from ours.}. In its bosonic representation, a similar model has been addressed in Refs.~\cite{Hartmann_2017, PhysRevE.97.020202}: its quantum properties have been considered at the level of the density matrix and related to the classical chaos in the thermodynamic limit.
It is even possible to characterize the environment inducing dissipation as a device performing random measurements on a pure state. In such perspective, one can look at the properties of quantum states along the single quantum trajectories, thus manipulating a stochastically evolving state vector~\cite{Plenio, Daley2014,J_Iwaniszewski_1995,N_Gisin_1992}, rather than solving the full master equation for the reduced density matrix. 
In the context of dissipative quantum chaos, the trajectory approach has been already employed on a single-degree-of-freedom quantum system~\cite{PhysRevLett.94.164101, PhysRevLett.95.164101}.  

Here we address a rather different scenario, since, in the thermodynamic limit, our model is exactly described by a classical mean-field dynamics, while at finite size it is a quantum many-body spin system. We study the properties of finite-size quantum states along the trajectories and compare them with the corresponding infinite-size classical mean-field dynamics.
We focus on the behavior of quantities assessing the many-body quantum \rep{coherence}{complexity}, such as the nonstabilizerness and the entanglement entropy. 

The nonstabilizerness, also called the {\it magic}, has been recently introduced as a quantifier of complexity of a quantum state, complementary to the entanglement~\cite{gottesman1998,gottesman1998-2, bravyi2005, howard2014, veitch2014, chitambar2019, seddon2019, zhou2020, liu2022,leone2022, Oliviero_2022, PhysRevA.107.022429, Rattacaso_2023, Odavic_2023, Leone_2024,PhysRevResearch.6.L042030,PRXQuantum.5.030332,russomanno2025efficientevaluationnonstabilizernessunitary}. It measures the distance from the set of stabilizer states, i.\,e., ``easy'' states which can be prepared using a specific class of quantum circuits that are efficiently simulatable with classical algorithms (the Clifford circuits)~\cite{veitch2014,howard2014-2,bravyi2016,bravyi2016-2,bravyi2019,heinrich2019,wang2019,wang2020,heimendahl2021,jiang2023,haug2023-3}. We compute it through the stabilizer R\'enyi entropy~\cite{leone2022, Oliviero_2022} averaged over
quantum trajectories and using the methods of Ref.~\cite{passarelli2024nonstabilizerness} for a permutationally invariant system. 

We find that the magic reaches an asymptotic long-time value displaying a scaling behavior with the system size that is related to the properties of the classical mean-field dynamics. In correspondence of a mean-field chaotic dynamics and in the absence of dissipation, such scaling is logarithmic (i.\,e., logarithmic with the accessible Hilbert-space dimension). This numerical finding supports a theoretical statement recently put forward in Ref.~\cite{Leone2021quantumchaosis}. Adding dissipation, the scaling becomes power-law, with an exponent showing plateaus for the parameter ranges corresponding to a mean-field chaotic dynamics. The value of the exponent in these plateaus increases with the dissipation strength, while it approaches the zero value (i.\,e., near to logarithmic scaling) for small dissipation strengths. Thus, for small decay rate, the prediction of Ref.~\cite{Leone2021quantumchaosis} seems to be robust.

On different grounds, the entanglement entropy is known to play a crucial role in the study of many-body dissipative quantum systems evolving along stochastic trajectories.
In fact, a flourishing number of recent works has been focusing on local measurements (either discrete or continuous in time)
performed in monitored quantum 
circuits~\cite{Li2018, Chan2019, Skinner2019, Szyniszewski2019, Vasseur2021, Bao2021, Nahum2020, Chen2020,Li2019, Jian2020, Li2021, Szyniszewski2020, Turkeshi2020, Lunt2021, Sierant2022_B, Nahum2021, Zabalo2020, Sierant2022_A, Chiriaco2023, Klocke2023},
as well as in non-interacting~\cite{DeLuca2019,Nahum2020, Buchhold2021,Jian2022, Coppola2022, Fava2023, Poboiko2023, Jian2023, Merritt2023, Alberton2021, Turkeshi2021, Szyniszewski2022, Turkeshi2022, Piccitto2022, Piccitto2022e, Tirrito2022, Paviglianiti2023, chahine2023entanglement,Kells_2023}
and interacting~\cite{Lunt2020,Rossini2020, Tang2020, Fuji2020, Sierant2021, Doggen2022, Altland2022,passarelli2023postselectionfree,delmonte2024measurementinducedphasetransitionsmonitored} Hamiltonian systems.
Moreover, a deep connection between measurement-induced phases and the encoding/decoding properties of a quantum channel has been put forward~\cite{Gullans2020_A, Gullans2020_B, Loio2023, Choi2020, Bao2020, Bao2021_A,Fidkowski2021, Bao2021_B, Barratt2022_A,Dehgani2023, Kelly2022}.
Situations where the dynamics is only induced by random measurements of non-local
string operators have been also considered~\cite{Ippoliti2021, Sriram2022}.
In our system we find that, quite unusually, the asymptotic long-time entanglement entropy averaged over trajectories behaves nonmonotonically with the system size: it increases up to a critical size and then starts decreasing. The maximum does not develop any discontinuity or peculiarity at the transition between regular and chaotic behaviors in the classical limit. 
Thus the entanglement entropy appears to be insensitive to classical chaos in the thermodynamic limit.

Other properties of the state along the trajectories may provide further indication on the classical chaotic behavior. One of them is the variance, over and along the trajectories, of the expectation value of a magnetization component on the evolving state~\cite{passarelli2023postselectionfree}. 
In the presence of chaos, the dependence of this variance on the system size develops an upper concavity, while in case of regular dynamics it shows a lower concavity. Focusing on the behavior of the full distribution of this expectation~\cite{Tirrito2022,Russomanno2023_longrange,piccitto2024impact}, we see that the chaos transition corresponds to a discontinuity of its global maximum. 

Summarizing, the main message of this work is that, while the entanglement is rather insensitive to the dynamic behavior in the classical mean-field limit of the dissipative quantum kicked top, the magic seems to be a good indicator of the underlying classical chaos. 

The paper is organized as follows. In Sec.~\ref{model:sec} we introduce the model. In Sec.~\ref{mean:sec} we study its classical mean-field behavior reached in the thermodynamic limit, describing the chaotic and regular regimes by means of different classical chaos probes (more details can be found in Appendix~\ref{other:sec}). In Sec.~\ref{traj:sec} we study the properties of quantum states along trajectories at finite size, focusing on the nonstabilizerness (Sec.~\ref{nosta:sec}), the entanglement entropy (Sec.~\ref{ent_ent:sec}), and the statistics of quantum trajectories (Sec.~\ref{fluc:sec}). Finally, in Sec.~\ref{conc:sec} we draw our conclusions.

\section{Model} \label{model:sec}

We consider a quantum kicked-top model~\cite{haake1987classical}, described by the Hamiltonian
\begin{align}\label{eq:kicked-top}
    \frac{\hat H(t)}{J} &= -2h \, \hat S_x + \frac{2}{N} \, \hat S_z^2 + \frac{2K}{N} \, \hat S_z^2 \sum_n \delta(t - t_n) \notag \\ &\equiv \hat H_0 + \hat H_K(t),
\end{align}
where $\hat S_\alpha = \tfrac12 \sum_{i=1}^N \hat \sigma^\alpha_i$ are collective spin operators ($\alpha = x, y, z$), $N$ is the number of spins, while $h$ and
$K$ respectively denote the strength of the transverse
field and of the periodic kick. This ensemble of particles is described by the total spin $S = N/2$. 
In Eq.~\eqref{eq:kicked-top}, the term $\hat H_0 = -2h \, \hat S_x + (2/N)  \, \hat S_z^2 $ corresponds to the Lipkin-Meshkov-Glick (LMG) model~\cite{lipkin1965}, which is periodically kicked by $\hat H_K(t)$ at times  $t_n = n \, \tau$, where $\tau$ is the driving period. In the following we fix $\tau = 1$ and omit it from the discussion; we also set the energy scale by fixing $J = 1$ and work in units of $\hbar = 1$.

Since the Hamiltonian~\eqref{eq:kicked-top} is periodic, it is useful to define the corresponding Floquet operator
\begin{equation}\label{eq:floquet}
    \hat{U}_F(\tau) = \hat{U}_K e^{-i \, \hat H_0 \tau},
\end{equation}
where
\begin{equation}\label{eq:kick}
    \hat{U}_K = e^{-i \, 2 K \hat S_z^2 / N}
\end{equation}
is the unitary operator responsible for the kicks.
This choice of $\hat{U}_K$ is known to induce quantum chaos for large enough values of $K$, as one can see from the properties of the eigenvalues and eigenstates of $\hat{U}_F(\tau)$~\cite{haake1987classical,PhysRevB.95.214307}. Classical and quantum chaos with a different driving have been studied in Ref.~\cite{Russomanno_2015}. Quantum chaos strictly corresponds to classical chaos appearing in the mean-field limit. This limit is exact for $N\to\infty$, since the components of the spin magnetization, $\hat m_\alpha = \hat S_\alpha / S $, commute for large $N$: $[\hat m_\alpha, \hat m_\beta] = i\, \epsilon_{\alpha \beta\gamma} \, \hat m_\gamma / S \xrightarrow[]{N\to\infty} 0$,
with $\epsilon_{\alpha, \beta \gamma}$ being the Levi-Civita symbol. The classical limit can display both regular and chaotic dynamics, closely mirrored by the properties of $\hat{U}_F(\tau)$ at large $N$~\cite{haake1987classical}.

Our goal is to go beyond the unitary dynamics, adding dissipation to the above model. We study it by means of the Lindblad master equation
\begin{equation}\label{eq:lindblad}
    \partial_t \hat\rho = -i \, [\hat H(t), \hat \rho] + \frac{\gamma}{S} \Big( \hat S_- \hat \rho \hat S_+ - \tfrac{1}{2} \lbrace \hat S_+ \hat S_-, \hat \rho\rbrace \Big),
\end{equation}
where $\hat L = \sqrt{\gamma/S} \, \hat S_-$ is the jump operator and $\gamma$ the decay rate, renormalized by the factor $S$ to have a well-defined thermodynamic limit. The collective Lindblad operators are the same used in Refs.~\cite{passarelli2023postselectionfree, PhysRevLett.121.035301}, in a case without periodic driving.

\section{Mean-field theory} \label{mean:sec}

In this section, we review some known results of the semiclassical mean-field limit of the model under study. We use the methods of Refs.~\cite{PhysRevLett.121.035301,PhysRevB.106.224308}. We start by computing the Heisenberg equations of motion for the total spin components. First of all, we consider the dynamics between two subsequent kicks, where the Dirac delta in Eq.~\eqref{eq:kicked-top} is zero. We obtain the following equations, omitting the time dependence of the operators in the Heisenberg picture for brevity:
\begin{subequations}\label{eq:spin-dynamics}
\begin{align}
    \partial_t \hat S_x = & \frac{1}{S} \big( \hat S_z \hat S_y + \hat S_y \hat S_z \big) + \frac{\gamma}{S} \big( \hat S_z \hat S_x + \hat S_x \hat S_z - \hat S_x \big),\\
    \partial_t \hat S_y = & -\frac{1}{S} \big( \hat S_z \hat S_x + \hat S_x \hat S_z \big) + 2 h \, \hat S_z \nonumber \\
    &\qquad + \frac{\gamma}{S} \big( \hat S_z \hat S_y + \hat S_y \hat S_z - \hat S_y \big), \\
    \partial_t \hat S_z = &- 2 h \, \hat S_y - \frac{2\gamma}{S} \big( \hat S_x^2 + \hat S_y^2 \big).
\end{align}
\end{subequations}

\begin{figure*}[t]
    \centering
    \includegraphics[width = \textwidth]{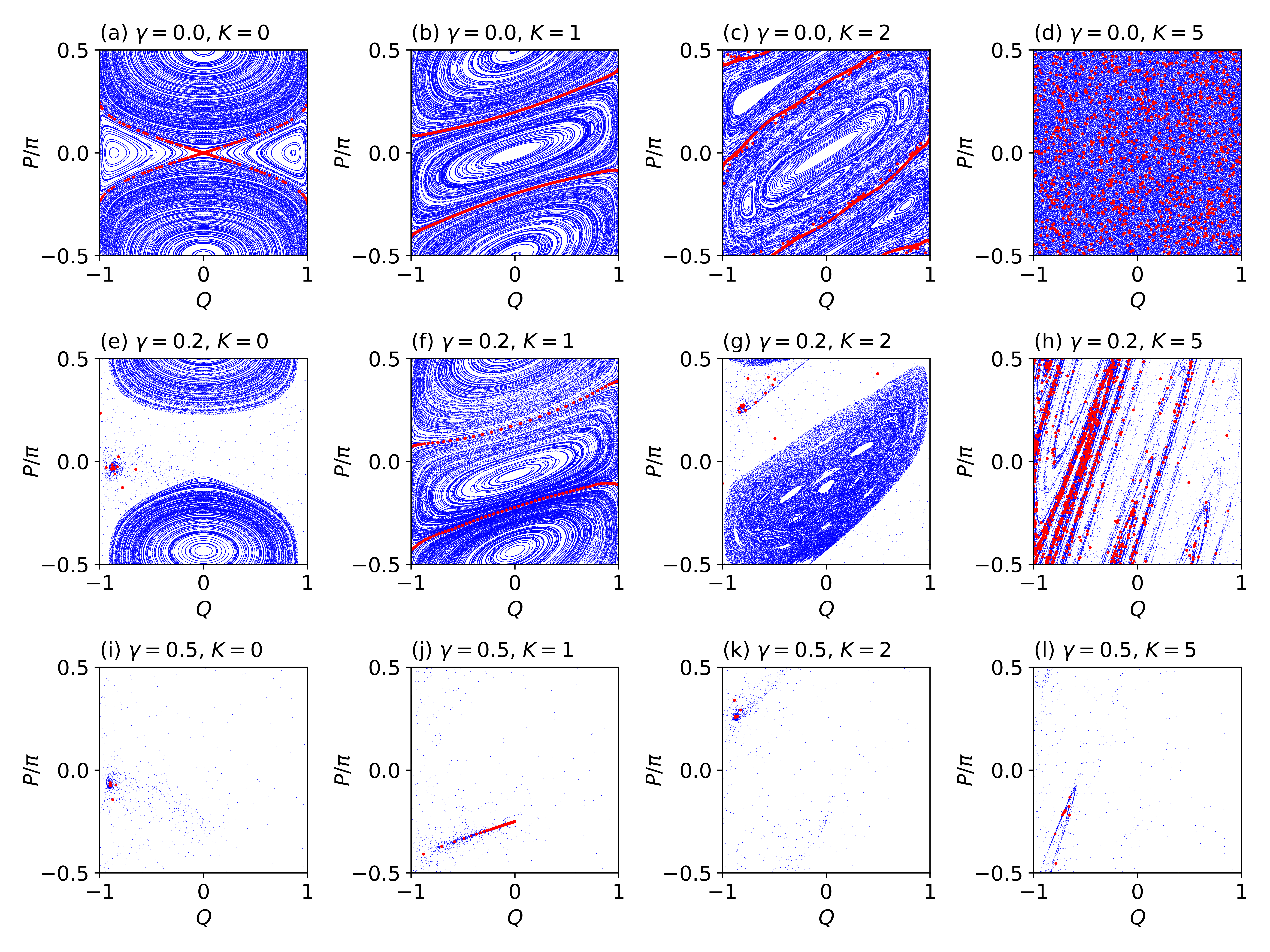}
    \caption{The Poincaré sections resulting from the mean-field equations for $h = 0.5$ and different choices of $K$ and $\gamma$ [see Eqs.~\eqref{eq:magnetization-dynamics} and~\eqref{eq:magnetization-kick}]. Red dots show the stroboscopic evolution of the initial state $(Q_0, P_0) = (1, 0)$, corresponding to $\vec{m}(0) = (0, 0, 1)$. Columns from left to right: $K=0, \, 1, \, 2, \, 5$. Rows from top to bottom: $\gamma=0, \, 0.2, \, 0.5$.}
    \label{fig:mf-poincare}
\end{figure*}

Recalling the definition of $\hat m_\alpha$ and defining $m_\alpha = \langle \hat m_\alpha \rangle $, we obtain the following equations, valid up to terms of order $O(1/N)$:
\begin{subequations}
\label{eq:magnetization-dynamics}
\begin{align}
    \partial_t m_x & = 2 \,m_y m_z + 2\gamma \,m_x m_z , \\
    \partial_t m_y & = -2 \,m_x m_z + 2h \, m_z + 2\gamma \,m_y m_z , \\
    \partial_t m_z & = - 2 h \, m_y - 2\gamma \left(m_x^2 + m_y^2\right).
\end{align}
\end{subequations}
We notice that, at the mean-field level, the norm $\lVert \vec{m} \rVert = \sqrt{\vec m \cdot \vec m}$ of $\vec{m} = (m_x, m_y, m_z)$ is conserved. Thus, we can always consider $\lVert \vec{m} \rVert = 1$. This allows us to \add{reduce the number of degrees of freedom from three to two and to} express the magnetization components in terms of just two variables, $Q$ and $P$, \add {with the change of variables}~\cite{Sciolla_2013}
\begin{subequations}
\begin{align}
    m_x & = \sqrt{1-Q^2} \cos 2P,\\ 
    m_y & = \sqrt{1-Q^2} \sin 2P, \\
    m_z & = Q,
\end{align}
\end{subequations}
where $Q \in [-1, 1]$ and $P \in [-\pi/2, \pi/2)$. The inverse relations are $Q = m_z$ and $P = \arctan(m_y / m_x)$, except when $Q = 1$, where one can consider $P = 0$. The variables $(Q, P)$ define the Poincaré section \add{and allow for a direct representation of the magnetization dynamics in a two-dimensional space (cf.~Fig.~\ref{fig:mf-poincare})}.

Secondly, we write down the equations describing the action of a kick on a given magnetization $\vec m$. To do so, we notice that, by the mean-field approximation, we can write
\begin{equation}
    \hat S_z^2 \approx S_z^2 + 2 S_z \, (\hat S_z - S_z) = 2 S_z \hat S_z - S_z^2,
\end{equation}
where $S_z$ denotes the expectation value of the operator $\hat S_z$. Thus, the unitary operator representing the action of the kick becomes
\begin{align}
    \hat U_K &= e^{-i \, K \hat S_z^2 / S} \approx e^{i K S_z^2 / S} e^{-2 i K S_z \hat S_z/ S} \notag \\ &\equiv e^{i \theta} \hat{\mathcal{R}}_z(2 K m_z),
\end{align}
where $\theta$ is an inessential phase and $\hat{\mathcal{R}}_z(\phi)$ is the rotation operator around the $z$ axis by an angle $\phi = 2 K m_z $. Therefore, at the mean-field level, the action of the kick is to change the magnetization $\vec m$ to $\vec{m}'$ such that
\begin{subequations}\label{eq:magnetization-kick}
\begin{align}
    m'_x & = m_x \cos(2 K m_z) - m_y \sin(2 K m_z),\\
    m'_y & = m_x \sin(2K m_z) + m_y \cos(2 K m_z),\\
    m'_z & = m_z.
\end{align}
\end{subequations}

\subsection{Phase portraits}

Equations~\eqref{eq:magnetization-dynamics} and~\eqref{eq:magnetization-kick} allow us to study the system dynamics at stroboscopic times (i.\,e., at integer multiples of $\tau$) in a way similar to Poincar\'e sections~\cite{Berry_regirr78:proceeding}\footnote{We consider a set of different initial configurations of values $(Q_0, P_0)$ that are randomly extracted from uniform distributions covering their respective domains. We consider $N_\tau = 10^3$ periods; in each period, we first evolve the magnetization using Eqs.~\eqref{eq:magnetization-dynamics}, then apply the kick following Eqs.~\eqref{eq:magnetization-kick}. We store the values $\boldsymbol{(}Q(t_n^+), P(t_n^+)\boldsymbol{)}$ right after the kick and use them to set the new initial condition for the following period. We repeat this simulation for $M_0 = 300$ initial configurations.}.
A selection of typical phase-space portraits is shown in Fig.~\ref{fig:mf-poincare}, for $h = 0.5$ and several choices of $K$ (different columns) and $\gamma$ (different rows). \add{In all panels, the red dots represent the trajectory of the initial state $\vec{m}(0) = (0, 0, 1)$, corresponding to $(Q_0, P_0) = (1, 0)$.} The first column corresponds to $K = 0$, that is, to the boundary time crystal model~\cite{PhysRevLett.121.035301}. In this case, when $\gamma = 0$, the initial states evolve following closed orbits and the dynamics is always regular. By increasing $\gamma$, we see that one of the fixed points becomes attractive and there is a coexistence of closed orbits and the stable fixed point. When the decay rate increases further [panel (i), for $\gamma=0.5$], closed orbits disappear and there is only one stable fixed point.

Moving along the next columns, we show results for increasing values of the kick strength $K$. When $K = 1$ (second column), the phase-space portrait still shows closed orbits and regular dynamics in the unitary limit ($\gamma=0$), and the appearance of a stable fixed point for larger $\gamma$. For larger values of $K$ phase portraits start showing chaotic regions. In particular, for intermediate values of the decay rate $\gamma$ (panel h) we observe strange attractors, that are a signature of dissipative chaos [the phase-space portrait for $K = 5$ in panel (h) is similar to the Zaslavski map~\cite{zasla}]. Stronger decay rates (bottom column) destroy these features and the resulting dynamics is always regular. However, there is a finite window of parameters in which the system displays peculiar chaotic behaviors as a result of the interplay between the driving (the kick) and the dissipation.

\subsection{Largest Lyapunov exponent $\lambda$}

\begin{figure}[!t]
    \centering
    \includegraphics[width = \columnwidth]{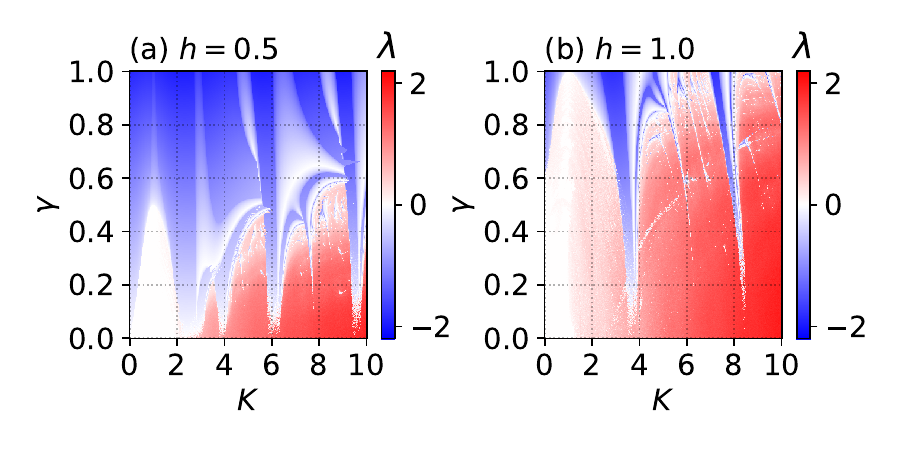}
    \caption{The largest Lyapunov exponent $\lambda$ as a function of the kick strength $K$ and the decay rate $\gamma$. (a), left panel: $h = 0.5$. (b), right panel: $h = 1.0$. The value $h = 0.5$ corresponds to the Poincaré sections shown in Fig.~\ref{fig:mf-poincare}.}
    \label{fig:lyapunov-exponent}
\end{figure}

To make more quantitative statements, we evaluate the largest Lyapunov exponent $\lambda$, which measures the exponential divergence of nearby trajectories~\cite{Piko}. Given two trajectories whose initial distance is $d_0 = \lvert \vec{m}_1(0) - \vec{m}_2(0)\rvert$, this is defined as
\begin{equation}
\lambda = \lim_{d_0\to 0} \, \lim_{t\to\infty} \ln \frac{d(t)}{d_0},
\end{equation}
where $d(t)$ is the distance between the trajectories at time $t$ and $d_0\equiv d(0)$\footnote{We evaluate $\lambda$ as explained in Ref.~\cite{PhysRevA.14.2338}. More precisely, we start from two nearby initial conditions at distance $d_0$, evolve them for one period and measure their distance $d_1$. Then we move one of the two points along the line joining them to restore the distance to the value $d_0$ and repeat, at each period $k$ storing the distance between the final points of the trajectories $d_k$. The largest Lyapunov exponent is then estimated as $\lambda = (1 / N_\tau \tau) \sum_{k=1}^{N_\tau} \ln(d_k / d_0) $. We consider the two initial conditions $\vec{m}_1(0) = (0, 0, 1)$ and $\vec{m}_2(0) = (\sqrt{d_0/2}, \sqrt{d_0/2}, \sqrt{1-d_0^2})$ with $d_0 = 10^{-10}$, evolving the system for $N_\tau = 1000$ periods.}.
We scan the parameter intervals $\gamma \in [0, 1]$ and $K \in [0, 10]$ (using a grid of $400 \times 400$ points). Our results are shown in the heat maps of Fig.~\ref{fig:lyapunov-exponent} for two values of the transverse field $h$. White ($\lambda=0$) and blue ($\lambda<0$) colors mark the regions where the dynamics is regular. More precisely, white regions correspond to the existence of closed orbits in the Poincaré sections of Fig.~\ref{fig:mf-poincare}, whereas blue regions are a footprint of the existence of a stable fixed point or cycle in the dynamics. 

In contrast, red regions ($\lambda>0$) are chaotic: these show up when the kick strength is sufficiently large and tends to disappear by increasing the decay rate $\gamma$. Nonetheless, there exist regions of this phase diagram where the chaotic behavior survives even for finite decay rates, in agreement with our previous observations. We also notice that chaotic regions become prominent for larger values of $h$ (compare the right panel of Fig.~\ref{fig:lyapunov-exponent} with the left one), in the sense that, for a given kick strength $K$, a stronger decay rate $\gamma$ is needed to regularize the dynamics. Apart from this quantitative difference, we do not observe any other significant dependence on the value of $h$, thus, unless differently specified, in the following we always focus on $h = 0.5$ (left panel).
We remark that, for this choice of $h$, chaotic regions can extend up to $\gamma \approx 0.6$; above this threshold, the dynamics is always regular for all the considered values of $K$.

The analysis of other probes for classical chaos (such as the bifurcation plot and the Hausdorff dimension) is reported in Appendix~\ref{other:sec}. In the following, we rather concentrate on the description of the behavior of the quantum system at finite size along the stochastic quantum trajectories, which constitutes the main part of this work.

\section{Finite-size dynamics along quantum trajectories}\label{traj:sec}

We now consider the same system at finite size, where the dynamics is genuinely quantum. Our goal is to see whether the different dynamic regimes emerging in the (classical) thermodynamic limit (cf.~Sec.~\ref{mean:sec}) reflect on the behavior of some functions of the quantum states\footnote{\textit{En passant}, we recall that expectations of collective magnetization components follow the mean-field values up to a time that scales with $N$. The scaling is different in the two regimes, being $O(N)$ for a regular dynamics and $O(\ln N)$ for a chaotic one~\cite{Schubert_2012, PhysRevB.98.134303}.}. 
Since the Hamiltonian~\eqref{eq:kicked-top} commutes with the total spin operator $\hat S^2 = \hat{\vec{S}} \cdot \hat{\vec{S}}$ and the Lindblad operator preserves this symmetry, we can restrict our attention to the permutationally invariant (maximum spin) sector, where $S = N / 2$ and the Hilbert space dimension is $D = N + 1$. In fact, the symmetry under permutations allows us to work in the subspace spanned by the Dicke states, defined as
\begin{equation}
    \Ket{\frac{N}{2}, \frac{N}{2} - k} = {\binom{N}{k}}^{-1/2} \, \hat{\mathcal{S}}\left(\ket{\uparrow}^{\otimes (N - k)} \otimes \ket{\downarrow}^{\otimes k}\right),
    \label{eq:Dicke}
\end{equation}
where $\hat{\mathcal{S}}$ is the symmetrization operator and $k \in \lbrace 0, 1, \dots, N \rbrace $.

As probes of quantum chaos, we use two information-theoretic quantities, (i) the stabilizer 2-R\'enyi entropy (SRE) $\mathcal{M}_2$ and (ii) the von Neumann bipartite entanglement entropy (EE) $\mathcal{E}$. 
We evaluate both quantities for pure states along the quantum trajectories, randomly generated in such a way that, on average, they describe the dynamics of the Lindblad equation [Eq.~\eqref{eq:lindblad}]. In particular, we adopt a quantum-jump unraveling~\cite{Plenio} of the Lindbladian. This unraveling protocol works as follows.

First of all, the kicks act at times $t = t_n$, by applying to the quantum state the operator $\hat{U}_K$ defined in Eq.~\eqref{eq:kick}. Between two consecutive kicks, the quantum-jump protocol is the same used in Ref.~\cite{passarelli2023postselectionfree}. We discretize the time in intervals of length $\delta t$. Here the unraveling works by defining the non-Hermitian (nH) Hamiltonian
\begin{equation}\label{eq:effective-hamiltonian}
    \hat H_\text{nH} = \hat H_0 - i \, \frac{\gamma}{2S} \, \hat S_+ \hat S_-\,.
\end{equation}
At each infinitesimal time step $\delta t$, the state $\ket{\psi_t}$ can either undergo a ``no-jump'' evolution generated by $\hat H_\text{nH}$, with probability $1 - (\gamma/S) \braket{\psi_t|\hat S_+ \hat S_-|\psi_t} \delta t $, or it can instantaneously ``jump'' to 
\begin{equation}
    \ket{\psi(t + \delta t)} = \frac{\hat S_- \ket{\psi(t)}}{\lVert \hat S_- \ket{\psi(t)} \rVert}\,.
\end{equation}
Averaging over randomness and performing the limit $\delta t\to 0$, one recovers the Lindblad equation Eq.~\eqref{eq:lindblad}~\cite{Plenio, Daley2014}. 
In particular, the average state of an ensemble of $N_\text{traj}$ quantum trajectories characterized by pure states $\ket{\psi_k}$ is given by
\begin{equation}
    \hat \rho = \lim_{N_\text{traj}\to\infty}\frac{1}{N_\text{traj}} \sum_{k = 1}^{N_\text{traj}} \ket{\psi_k}\!\bra{\psi_k}.
\end{equation}
The expectation value of any operator $\hat O$, which we denote by angular brackets $\langle \hat O \rangle$, can be computed by averaging its expectation values over the individual trajectories:
\begin{equation}
    \mathrm{Tr}(\hat \rho \, \hat O) = \lim_{N_\text{traj}\to\infty}\frac{1}{N_\text{traj}} \sum_{k=1}^{N_\text{traj}} \braket{\psi_k | \hat O | \psi_k} \equiv \langle \hat O \rangle.
    \label{eq:avg_traj}
\end{equation}

While averaging over pure-state trajectories is not essential to get expectation values of observables, this procedure becomes crucial whenever evaluating a nonlinear function of the quantum state, such as for the SRE or the EE (their definitions are provided in the following subsections). For instance, the bipartite von Neumann entropy of the average state $\hat \rho$ is not a good entanglement indicator, as it includes classical correlations as well. \rep{On the other hand, thanks to the unraveling into pure-state trajectories, it is possible to define the average entanglement of the ensemble as
$\langle \mathcal{E} \rangle = \frac{1}{N_\text{traj}} \sum_{k=1}^{N_\text{traj}} \mathcal{E}(\ket{\psi_k}) \ne \mathcal{E}(\hat \rho)$,
and similarly goes for the average magic $\langle \mathcal{M}_2 \rangle$.}{In fact, given a generic nonlinear functional $\mathcal{F}$ of $\hat \rho$, such as $\mathcal{E}$ or $\mathcal{M}_2$, one can define its average value over the ensemble of pure-state trajectories as
\begin{equation}\label{lag:eqn}
    \langle \mathcal{F} \rangle = \frac{1}{N_\text{traj}} \sum_{k=1}^{N_\text{traj}} \mathcal{F}(\ket{\psi_k}\!\bra{\psi_k}), 
\end{equation}
which, in general, does not coincide with the evaluation of $\mathcal{F}$ on
the state $\hat \rho$ itself.
That is, contrary to expectations of observables, $\langle \mathcal{F} \rangle \ne \mathcal{F}(\hat \rho)$.}

From a numerical point of view, one is forced to consider a finite number of trajectories $N_\text{traj}$. This gives rise to an error bar in the average value that is the root mean square deviation divided by $\sqrt{N_\text{traj}}$. In the following, when studying dissipative evolutions, we fix $N_\text{traj} = 1024$, in such a way that the resulting error bars are always of the order of the line width of our plots. 
As for the initial state, we consider the fully polarized eigenstate $\ket{\psi(0)} = \ket{N/2, N/2}$ of the $\hat S_z$ total spin component.

Sometimes we are also interested in the asymptotic (long-time) values of such quantities. We estimate them by evolving the system up to long times, until a plateau is reached. In our investigations we find that this saturation is achieved quickly, so that we can stop the dynamics after $N_\tau = 1000$ periods for the entanglement entropy and $N_\tau = 100$ for the magic. The value at the plateau always fluctuates, due to the fact that we are considering a finite ensemble of quantum trajectories. Thus, to extract the steady-state behavior, we average over the last $k_0$ points, with $k_0$ large enough that all transient features have disappeared (specifically, we set $k_0 = 500$ for the EE and $k_0 = 50$ for the SRE). We denote this long-time average by $\overline{\langle \dots \rangle}$ such that, \add{given any nonlinear functional $\mathcal{F}$,}
\begin{equation}\label{gen:eqn}
    \overline{\langle {\add{\mathcal{F}}}\rangle} = \frac{1}{N_\tau-k_0}\sum_{k=k_0}^{N_\tau} \langle \add{\mathcal{F}} (t_k)\rangle,
\end{equation}
\add{where $\langle \mathcal{F} (t_k)\rangle$ is the average of $\mathcal{F}$ over the states of the ensemble at time $t_k$.} 
We also compare dissipative evolutions ($\gamma > 0$) with the unitary dynamics case ($\gamma = 0$). Unitary results are averaged over $M_0 = 300$ initial random spin coherent states~\cite{PhysRevA.6.2211}. In fact, the dynamics of a single trajectory entails a source of randomness (the jumps) that is absent in the unitary case. Thus, in the presence of dissipation, the system is able to explore more efficiently the Hilbert space and there is no need for averaging over the different initial conditions.

\subsection{Nonstabilizerness}\label{nosta:sec}

We now focus on the connection between chaos and nonstabilizerness.
To define the latter, we first consider the Clifford group $\mathcal{C}_N$: %
this is spanned by the single-qubit Hadamard ($\mathsf{H}$) and phase ($\mathsf{S}$) operators, and the two-qubit $\mathsf{CNOT}$ gate. In quantum computation, universality is obtained by adding the $\mathsf{T}$-phase gate to the above Clifford group. Nonstabilizerness, or quantum magic, expresses how far the state is from the set of stabilizers, i.\,e., the set of states that can be prepared by only using quantum gates from $\mathcal{C}_N$. There exist efficient classical algorithms to manipulate stabilizer states~\cite{gottesman1997,aaronson2004}, whereas states beyond the stabilizer formalism are hard to study classically. Thus, when combined with the right amount of entanglement, nonstabilizer states are a necessary ingredient to unlock quantum advantage~\cite{beverland2020,wang2019}.

In recent years, several measures of nonstabilizerness have been proposed~\cite{veitch2014,howard2014-2,bravyi2016,bravyi2016-2,bravyi2019,heinrich2019,wang2019,wang2020,heimendahl2021,jiang2023,haug2023-3,beverland2020,turkeshi2023,tirrito2023}. Among them, we choose the SRE of the Pauli spectrum, defined as~\cite{leone2022}
\begin{equation}\label{eq:sre}
    \mathcal{M}_2(\ket{\psi}\add{\!\bra{\psi}}) = -\ln \Biggl[\sum_{\hat P\in\mathcal{P}_N} \frac{\braket{\psi | \hat P | \psi}^4}{2^{2N}}\Biggr] - N \ln 2,
\end{equation}
where $\hat P = \hat \sigma_1^{\alpha_1} \otimes \dots \otimes \hat\sigma_N^{\alpha_N}$ (with $\alpha_j \in \lbrace 0, x, y, z \rbrace$) are the Pauli strings and $\mathcal{P}_N$ is the Pauli group. The SRE is zero for stabilizer states and positive otherwise; it is generally an extensive quantity and is upper-bounded by $N \ln 2$. Evaluating it for general systems is exponentially hard, since both the state dimension and the number of Pauli strings grow exponentially with the number $N$ of two-level systems. Luckily, the permutation symmetry yields an exponential advantage in this respect, since it exploits the fact that Pauli strings differing by permutations have the same expectation values on a permutationally-symmetric quantum state, thus allowing us to study systems otherwise out of reach for numerical analysis~\cite{passarelli2024nonstabilizerness}.

\begin{figure}[tb]
    \centering
    \includegraphics[width=\columnwidth]{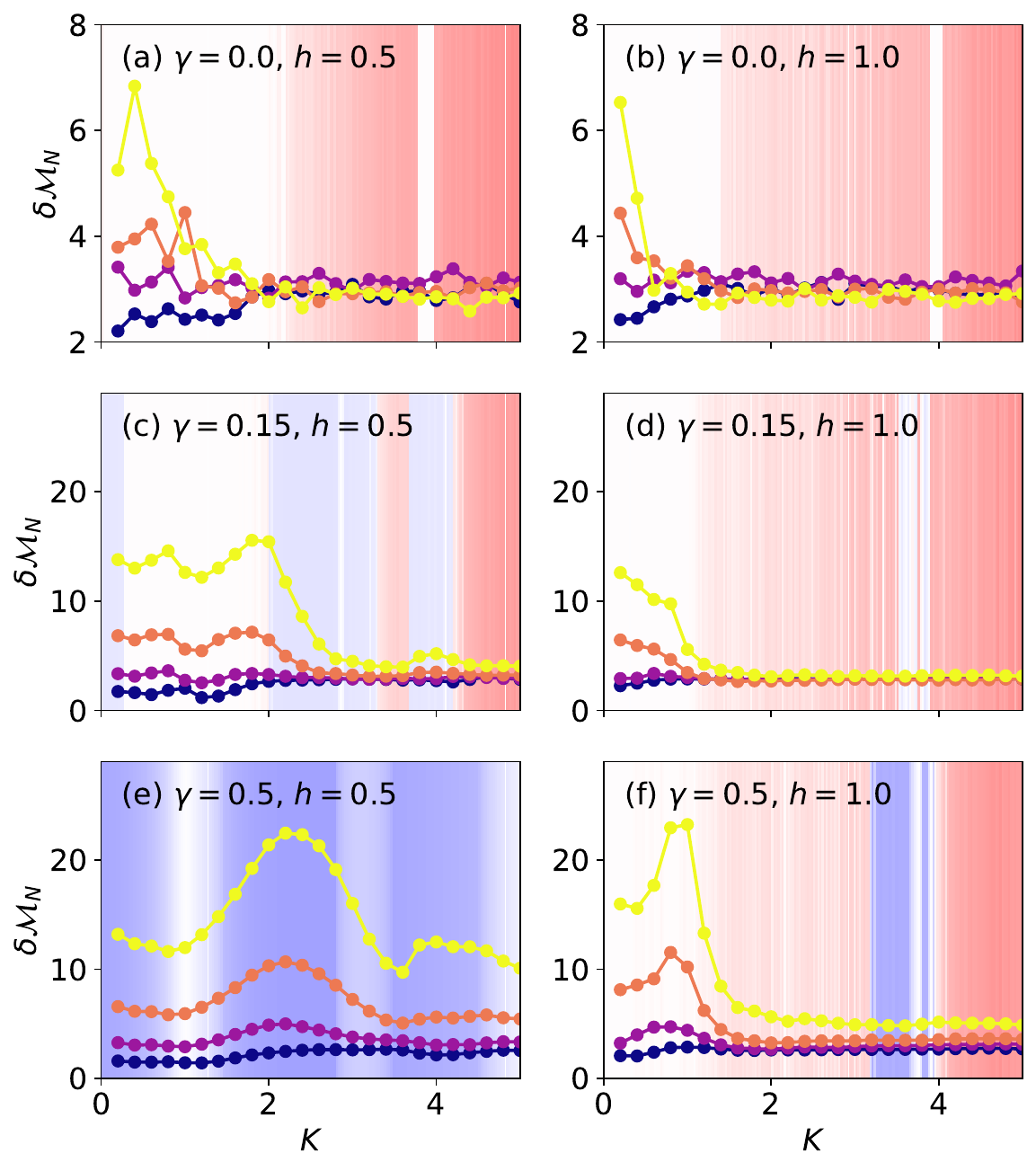}
    \caption{The quantity $\delta \mathcal{M}_N$ versus the kick strength $K$, for several combinations of $\gamma$, $h$ (see the headers of the panels). The various curves, with different color scale, refer to $N \in \lbrace 20, 40, 80, 160\rbrace$,\rep{ following the same color scheme as in Fig.~\ref{fig:sre-dynamics}}{ from $N = 20$ (darkest) to $N=160$ (lightest)}. %
    The background color is related to the value of the largest Lyapunov exponent $\lambda$, following the same color code as in Fig.~\ref{fig:lyapunov-exponent}.}
    \label{fig:sre-diff-scaling}
\end{figure}

\add{By looking at the stroboscopic evolution of $\langle \mathcal{M}_2\rangle $ averaged over the randomness, we notice that the dynamics converges to plateaus for all $N$ in a short time scale. Interestingly, the magic value at the plateaus seems to scale differently with the system size, depending on the value of $K$. When the kick strength $K$ is small and the dynamics is regular, the asymptotic value seems to grow extensively with $N$. On the other hand, for larger kick strengths, the dynamics becomes chaotic and the asymptotic value increases more slowly, possibly as $\ln N$. These data are shown in Appendix~\ref{app:numerics}.} To \rep{further investigate}{quantitatively clarify} this point, here we define 
the difference between the asymptotic SRE for a system of size $N$ and that for a system of size $N/2$:
\begin{equation}
    \delta \mathcal{M}_N = \overline{\langle \mathcal{M}_2\rangle}_{N} - \overline{\langle \mathcal{M}_2\rangle}_{N/2}\,.
\end{equation}
This quantity tends to a constant for increasing $N$, if the SRE increases logarithmically with $N$, while it keeps increasing for a 
super-logarithmic dependence of the SRE with $N$.
In Fig.~\ref{fig:sre-diff-scaling} we plot $\delta \mathcal{M}_N$ versus $K$, for different choices of the parameters $\gamma$, $h$ and for various sizes $N$. 
In the background of each panel, we also show the corresponding value of the largest Lyapunov exponent $\lambda$ in the classical limit, using the same chromatic scale adopted in Fig.~\ref{fig:lyapunov-exponent} (red regions are chaotic, blue regions are regular). In the classically chaotic regions and when $\gamma=0$ [panels (a) and (b)], we clearly see that $\delta \mathcal{M}_N$ saturates with $N$. This saturation persists, yet less distinctly, for $\gamma=0.15$ [panels (c), (d)] while it seems to fade away for $\gamma=0.5$ [panels (e), (f)] (although, in this latter case and for the chaotic regions we observe an increase with $N$, this is much slower than in the regular regions). \rep{We also see that, whatever the value of $\gamma$, in the classically chaotic regions $\delta \mathcal{M}_N$ is almost independent of $K$ (the same occurs for $\overline{\langle \mathcal{M}_2\rangle}_{N}$), in contrast with a strong dependence on $K$ that emerges in the classically regular regions.}{We also see that, in the classically chaotic regions, $\delta \mathcal{M}_N$ is almost independent of $K$ (the same occurs for $\overline{\langle \mathcal{M}_2\rangle}_{N}$, see Appendix~\ref{app:numerics}), in contrast with a strong dependence on $K$ that emerges in the classically regular regions. This is especially true for smaller values of $\gamma$, see panels (a-d).}\footnote{Results for $\gamma = 0$ are more noisy than those for $\gamma > 0$ since, as already mentioned, in the unitary case we average over $M_0 = 300$ initial spin-coherent states instead of the $N_\text{traj} = 1024$ trajectories of the dissipative case.}

\begin{figure}[!t]
    \centering
    \includegraphics[width = \columnwidth]{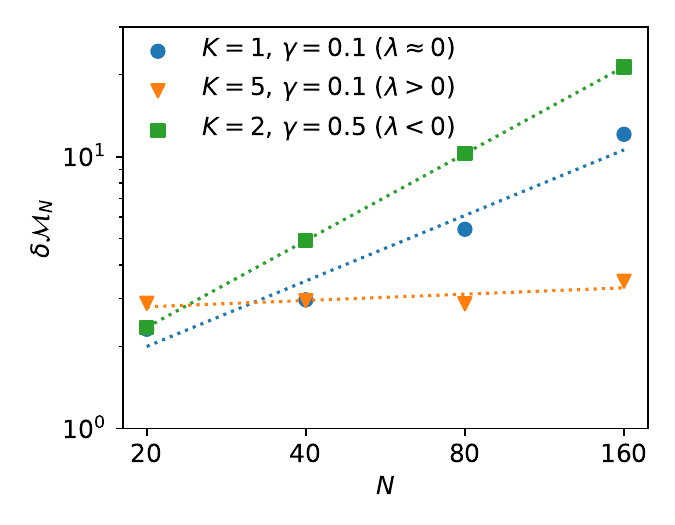}
    \caption{\del{Same plot as in Fig.~\ref{fig:sre-N}, but for }The quantity $\delta \mathcal{M}_N$ versus $N$. Dotted lines are the fitting functions $\delta \mathcal{M}_N \sim N^\beta$.}
    \label{fig:sre-diff-N}
\end{figure}

In Fig.~\ref{fig:sre-diff-N}, we plot $\delta \mathcal{M}_N$ versus the system size $N$, for \rep{the same}{some} choices of $K$, $\gamma$, and $h$\del{ of Fig.~\ref{fig:sre-N}}. We see that, for the parameters that correspond to a regular mean-field dynamics, $\delta \mathcal{M}_N$ strongly depends on $N$, whereas this dependence becomes much weaker when the dynamics is chaotic. We fit these curves with power-law functions $\delta \mathcal{M}_N \sim N^\beta$ (in this double-logarithmic scale, they appear as straight lines) and the values of $\beta$ resulting from the fit are reported in Tab.~\ref{tab:sre-scaling}\add{, together with the scaling exponents $\alpha$ of the fitting law $\overline{\langle \mathcal{M}_2 \rangle} \sim N^\alpha$.}

\begin{table}[tb]
    \centering
    \begin{tabular}{ccc}
    \toprule
       Parameters & $\alpha$ & $\beta$\\
       \midrule
        $K = 1$, $\gamma = 0.1$ & $0.89\pm 0.05$ & $0.80\pm0.13$ \\
        $K = 5$, $\gamma = 0.1$ & $0.359\pm0.003$ & $0.08\pm0.05$ \\
        $K = 2$, $\gamma = 0.5$ & $1.028\pm0.004$ &$1.061\pm0.002$ \\
        \bottomrule
    \end{tabular}
    \caption{Exponent of the scaling laws $\overline{\langle \mathcal{M}_2\rangle} \sim N^\alpha$ and $\delta \mathcal{M}_N \sim N^\beta$ for the parameters shown in \rep{Figs.~\ref{fig:sre-N} and}{Fig.}~\ref{fig:sre-diff-N}.}
    \label{tab:sre-scaling}
\end{table}

\begin{figure}[!t]
    \centering
    \includegraphics[width = \columnwidth]{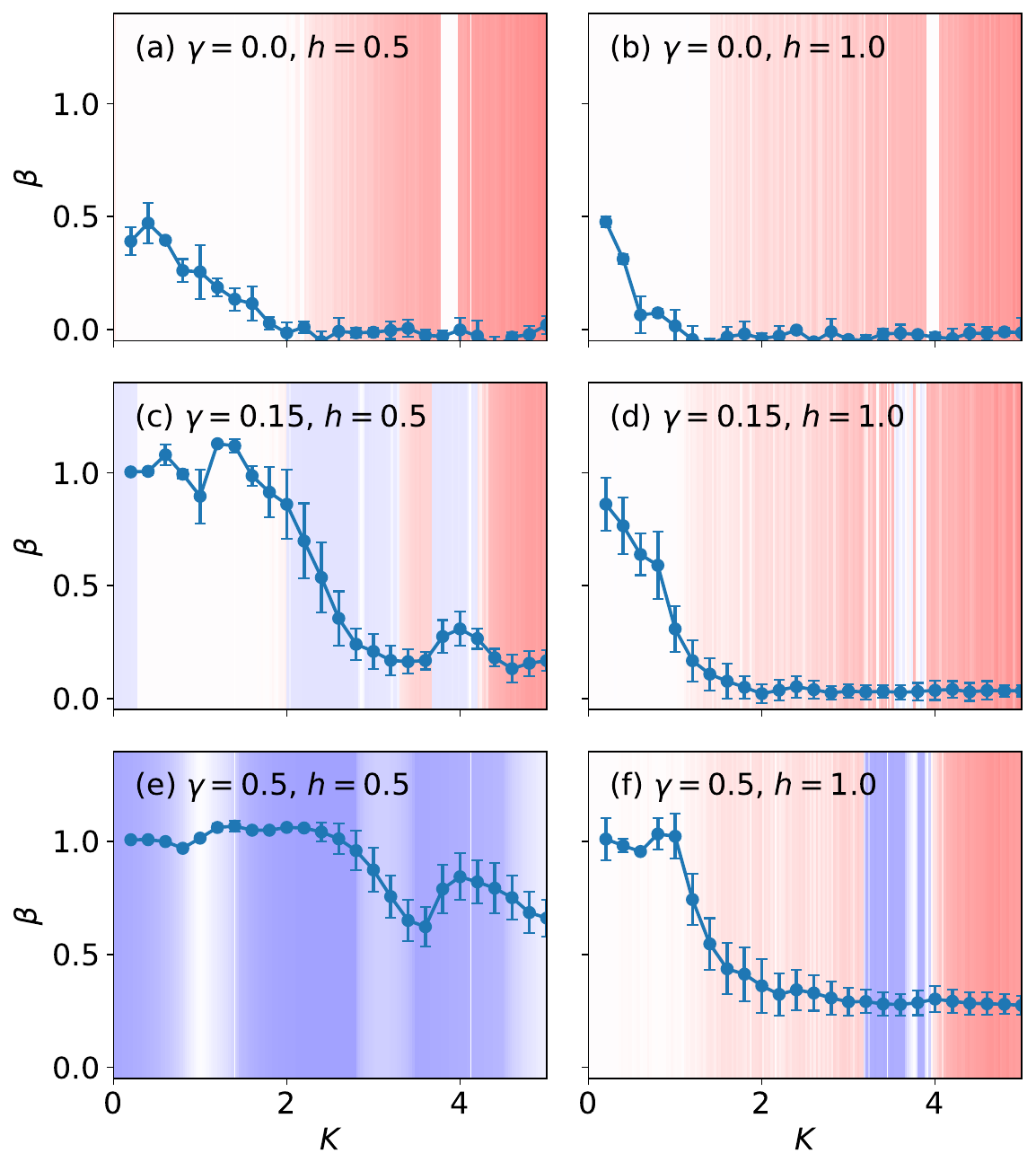}
    \caption{Exponent $\beta$ of the scaling law $\delta \mathcal{M}_N \sim N^\beta$ versus $K$, for different values of the decay rate $\gamma$ and transverse field $h$, see captions. The background color is related to the value of the largest Lyapunov exponent $\lambda$, following the same color code as in Fig.~\ref{fig:lyapunov-exponent}.}
    \label{fig:beta}
\end{figure}

In addition to this, Fig.~\ref{fig:beta} reports the behavior of $\beta$ versus $K$, for various and fixed values of $\gamma$, $h$. Let us first focus on the classically chaotic regions ($\lambda>0$, red background). We see that, for the unitary case [panels (a) and (b)], $\beta$ vanishes, meaning a logarithmic scaling of $\mathcal{M}_2$ with $N$\footnote{In the chaotic regime of the unitary case $\gamma = 0$, the exponent $\beta$ evaluated numerically only appears (slightly) negative due to noise in the curves shown in Fig.~\ref{fig:sre-diff-N}.}. Since the dimension of the available Hilbert space is $D = N + 1$, this implies $\mathcal{M}_2 \sim \ln D$. This finding confirms the prediction of Ref.~\cite{Leone2021quantumchaosis}. In that reference, a scaling $\mathcal{M}_2 \sim \ln D$ is shown to emerge for chaotic quantum systems under unitary dynamics, but the focus is on non-symmetric systems made up of $N$ two-level systems, so $D = 2^N$, \add{thus implying an extensive scaling of $\mathcal{M}_2$ with $N$.} \add{In contrast, in our case the dimension of the available Hilbert space is $D = N + 1$, thus we have $\mathcal{M}_2 \sim \ln N$. This observation may make stabilizer-based simulations of such
states more feasible, likely due to the unique properties of chaotic states in reduced-dimensional spaces.}

\add{At this stage, it is important to stress that permutational invariance does not inherently constrain the maximum achievable SRE, as $\mathcal{M}_2^\text{(max)} = N \ln 2$ could still be reached even with permutationally invariant states. This bound arises from a flat Pauli spectrum where each Pauli string has an exponentially small weight of $2^{-N}$ over the quantum state---a condition achievable in permutationally invariant states. In fact, in the regular regime, we observe that $\mathcal{M}_2$ can scale extensively with $N$ [see, e.g., Fig.~\ref{fig:beta}(e)], demonstrating that this scaling is allowed with permutational invariance as well.}
\add{That said, %
our results indicate that, in the chaotic regime, the $N$-scaling of its ensemble average over initial states becomes logarithmic in case of unitary dynamics with a classical chaotic limit [see Fig.~\ref{fig:beta}(a,b)].} 

Switching on the coupling to the environment  ($\gamma\neq 0$), we find $\beta\simeq 0$ when $\gamma\ll h$ [Fig.~\ref{fig:beta}(d)], otherwise it displays a positive plateau in $K$ which is significantly smaller than one. Notice that one gets $\delta \mathcal{M}_N \sim N^\beta$ with $\beta>0$ when also $\overline{\langle \mathcal{M}_2\rangle}_{N}\sim N^\beta$, so in these chaotic regions the SRE behaves subextensively. On the other hand, when the mean-field dynamics is regular, $\beta$ does not show plateaus, but rather a nontrivial dependence on $K$. In particular, it may increase and become close to one, meaning an extensive behavior of the SRE\footnote{Due to finite-size effects, in certain regions a superextensive scaling of the magic can be possible ($\beta > 1$). However, the bound $\mathcal{M}_2 < N \ln 2$ tells us that this apparent anomaly is expected to disappear at large $N$.}. In certain regions that are characterized by alternating regular and chaotic behaviors [see for instance the rightmost part of Fig.~\ref{fig:beta}(c)], the exponent $\beta$ can be sensitive to this pattern and develop a peak.

\subsection{Entanglement entropy}\label{ent_ent:sec}
We now study the bipartite EE~\cite{Nielsen}, defined as
\begin{equation}\label{eq:entanglement-entropy}
    \mathcal{E}\add{(\hat \rho)} = -\mathrm{Tr}(\hat\rho_A \ln \hat\rho_A),
\end{equation}
where $A$ and $B$ are equal disjoint bipartitions of the fully-connected model under study and $\hat\rho_A = \mathrm{Tr}_B(\hat\rho)$ is the reduced density matrix of subsystem $A$ alone. Due to permutational invariance, only the number of spins in each partition matters; hereafter we consider $N_A = N_B = N/2$. We carefully verified that our results are robust to changes in the bipartition size.

\del{In Fig.~\ref{fig:entropy-dynamics} we show some examples of the EE $\braket{\mathcal{E}(t_n)}$, averaged over the trajectories, versus the stroboscopic time $t_n$.
The top row corresponds to the unitary limit $\gamma = 0$, for the two values of the kick strength $K = 0$ where the mean-field dynamics is regular [panel (a)], and $K = 5$ where it is chaotic [(panel (b)]. All the entropies increase and then saturate to the asymptotic values.
To be more quantitative, it is useful to focus on the properties of the asymptotic value $\overline{\langle \mathcal{E} \rangle}$. In the ergodic regime, we find $\overline{\langle \mathcal{E} \rangle} = \alpha + \beta \ln (N + 1)$, with $\alpha = -1.05\pm 0.07$ and $\beta = 0.974\pm 0.016$ for $K=5$. The fact that $\beta \approx 1$ agrees with the expectation that the EE reaches its maximum value $\mathcal{E}_\text{max} = \ln D$ (where $D=N+1$),
for an ergodic dynamics. In contrast, when $K$ is smaller, we observe a slower growth with the system size: using the same fitting function, for $K=0$ we get $\alpha = -0.17\pm 0.07$ and $\beta = 0.477 \pm 0.016$.}

\begin{figure}[t]
    \centering
    \includegraphics[width = \columnwidth]{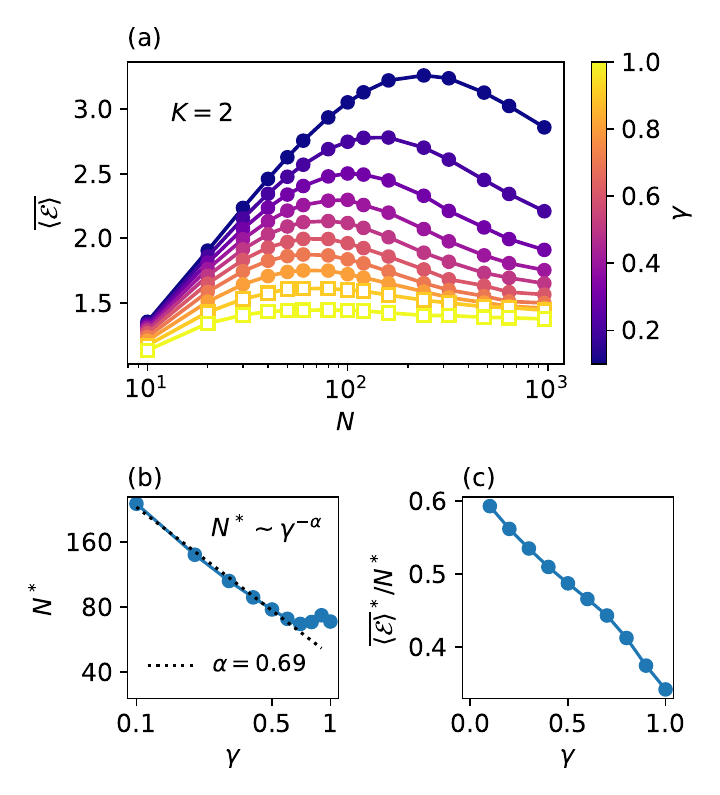}
    \caption{(a) $\overline{\langle \mathcal{E} \rangle}$ versus $N$ for several values of $\gamma$ and for fixed $h = 1$, $K = 2$. The curves refer to different values of $\gamma_j = j \, \delta \gamma$ with $\delta \gamma = 0.1$, from $j = 1$ (darkest) to $j= 10$ (lightest). The largest Lyapunov exponent $\lambda$ is positive for $\gamma \lesssim 0.87$ \add{(curves with filled circles)} and negative otherwise \add{(curves with empty squares)}. (b) Position $N^*$ of the maximum of $\overline{\langle \mathcal{E} \rangle}$ in panel (a) versus $\gamma$, log-log scale. (c) Maximum $\overline{\langle \mathcal{E} \rangle}$ divided by $N^*$ versus $\gamma$, linear scale.}
    \label{fig:entropy-scaling}
\end{figure}

\del{In the central and lower rows of Fig.~\ref{fig:entropy-dynamics}, we show the EE versus time in the presence of dissipation, being it either weak ($\gamma = 0.2$, middle row) or strong ($\gamma = 0.5$, bottom row), for the same values of $K$ considered before. Panels (c),(e), and (f) correspond to a regular mean-field dynamics ($\lambda<0$), while panel (d) is for a chaotic dynamics ($\lambda>0$). We observe that, after a transient, all curves converge to an asymptotic value.}

\add{As done in the previous section, we studied the entanglement dynamics, averaged over the trajectories, as a function of the stroboscopic time $t_n$. In all cases, we see that the entropy increases and then saturates to an asymptotic value (see Appendix~\ref{app:numerics}).}
Further insight can be obtained by focusing on the dependence of the asymptotic value $\overline{\langle \mathcal{E} \rangle}$ on $K$ and on the system size $N$. \add{This is reported in Appendix~\ref{app:numerics}, where we show that, by increasing $N$, the values of $\overline{\langle \mathcal{E} \rangle}$ display a tendency to saturate to a finite limit (area-law behavior) and even to decrease.}\del{In Fig.~\ref{fig:entropy-kick-scaling} we show the behavior of $\overline{\langle \mathcal{E} \rangle}$ versus $K$, for the same values of $\gamma$ and system sizes used in Fig.~\ref{fig:entropy-dynamics}. First of all, we notice a weak relation between such curves and the sign of the largest Lyapunov exponent $\lambda$ (see the color code in the background, as in Fig.~\ref{fig:sre-diff-scaling}). Furthermore, for increasing $N$, the values of $\overline{\langle \mathcal{E} \rangle}$ display a tendency to saturate to a finite limit (area-law behavior) and even to decrease.} To better understand this point, in Fig.~\ref{fig:entropy-scaling} panel (a) we report $\overline{\langle \mathcal{E} \rangle}$ versus $N$, for fixed $K$ and different values of $\gamma$. As a result of the competition between driving and dissipation, we observe a nonmonotonic behavior with the size $N$. 
We fix $K$ and $h$ because, with this choice of parameters, the maximum can be observed more easily, but this is a general phenomenon.
In fact, the nonmonotonicity occurs both for an underlying classical chaotic (curves with circles) and a regular (curves with squares) regime. However, our data suggest that all the curves eventually follow an area-law behavior, where $\overline{\langle \mathcal{E} \rangle}$ becomes independent of $N$.

Let us look more carefully to the maximum in $N$ of $\overline{\langle \mathcal{E} \rangle}$. The position $N^*$ of such maximum is plotted versus $\gamma$ in Fig.~\ref{fig:entropy-scaling}(b). For small decay rates, it follows a power-law scaling $N^* \sim \gamma^{-\alpha}$ with $\alpha \approx 0.69$, to eventually saturates when $\gamma$ is close to one. Saturation and transition of the mean-field dynamics from chaotic to regular are uncorrelated. In fact, even the maximum value ${\overline{\langle \mathcal{E} \rangle}}^*$ over $N^*$ keeps linearly decreasing with $\gamma$, irrespective of the dynamic features of the classical limit [Fig.~\ref{fig:entropy-scaling}(c)].
In summary, the asymptotic EE saturates towards an area-law behavior in a non-monotonous way, showing a maximum for a given system size. We are not able to find any relation between EE and chaos/regularity properties of the mean-field dynamics.

\subsection{Magnetization fluctuations}\label{fluc:sec}

\begin{figure}[!t]
    \centering
    \includegraphics[width=\columnwidth]{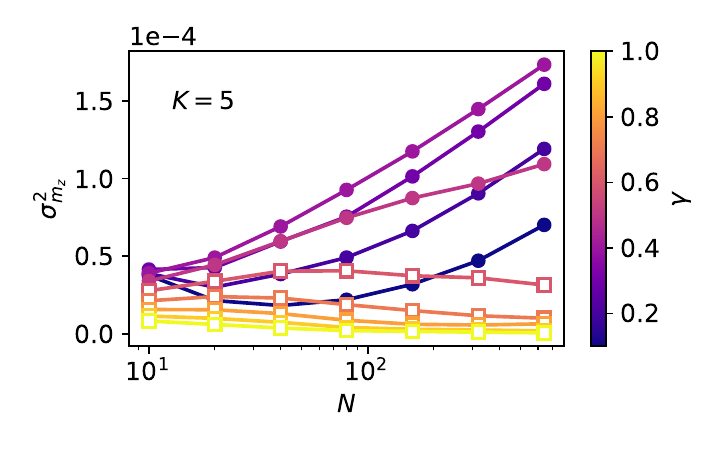}
    \caption{Variance of $m_z$ over time and randomness [Eq.~\eqref{osig:eqn}] as a function of the size, up to $N = 640$. The largest Lyapunov exponent is positive for $\gamma \lesssim 0.5$ \add{(curves with filled circles)} and negative otherwise \add{(curves with empty squares)}, as shown in Fig.~\ref{fig:lyapunov-exponent}(a).
    We fix $K = 5$ and $h = 0.5$.}
    \label{fig:variance}
\end{figure}

Another quantity of interest, related to chaos, is the variance, along and across the trajectories, of the magnetization component $m_z$ over the ensemble of quantum trajectories~\cite{passarelli2023postselectionfree}, defined as 
\begin{equation}\label{osig:eqn}
  \sigma_{m_z}^2 = \overline{\braket{m_z^2}}-\overline{\braket{m_z}}^2\,,
\end{equation}
where the average is over trajectories and time in the steady state, as defined in Eqs.~\eqref{lag:eqn} and~\eqref{gen:eqn}.
We fix the kick strength and study the scaling of these fluctuations as a function of the system size, for different decay rates. We show some examples thereof in Fig.~\ref{fig:variance}. We can distinguish two different behaviors for curves in which the mean-field dynamics is chaotic ($\lambda > 0$, circles) and those in which it is regular ($\lambda < 0$, squares). In the former case, curves start with a positive curvature and steadily increase with $N$, while in the latter case they start with a negative curvature and then tend to settle down to a small value.

\begin{figure}[!t]
    \centering
    \includegraphics[width=\columnwidth]{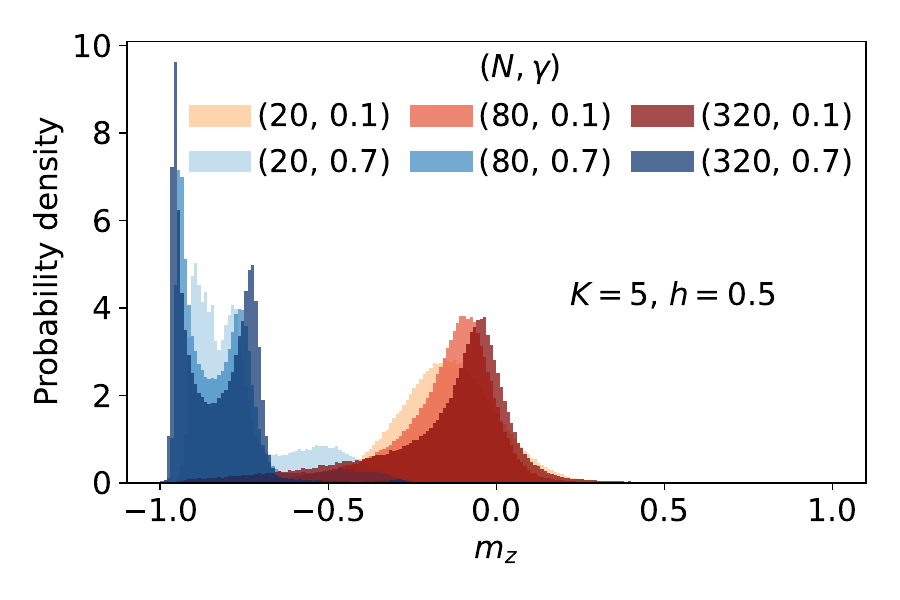}
    \caption{Probability density distributions of $m_z$ over 1024 trajectories and $200$ time points in the stationary state, for different values of $N$ and $\gamma$ (see legend). We fix $K = 5$ and $h = 0.5$.
    Red-like distributions correspond to a classically chaotic dynamics, while blue-like distributions are for a classically regular dynamics.}
    \label{fig:dist}
\end{figure}

To better understand this behavior, we have also considered the full distribution of $m_z(t_n)$ over randomness and stroboscopic time, similarly to what has been done in the absence of periodic driving~\cite{Tirrito2022, Russomanno2023_longrange, piccitto2024impact}. We obtain the distributions as normalized hystograms with bin size $\delta m_z = 0.01$, as reported in Fig.~\ref{fig:dist}. Here we fix $K=5$ and consider the two values of $\gamma= 0.1$ (chaotic mean-field dynamics) and of $\gamma=0.7$ (regular mean-field dynamics, with $\lambda<0$).
In the chaotic case (red-like histograms), the probability distribution shows a single peak located around a value close to zero. Increasing the system size, the distribution becomes skewed and develops a long tail towards negative values of the magnetization. This tail explains why, in the regime $\lambda > 0$, the variance in Fig.~\ref{fig:variance} increases with $N$.
Conversely, in the regular regime (blue-like histograms), the probability distribution is bimodal and values become closer to $m_z = -1$. For small $N$ we also observe a third peak around $m_z = -0.5$, which tends to disappear at larger $N$, where the distribution converges to a purely bimodal shape. This finding is consistent with the fact that, in this regime, the variance only depends on the system size when $N$ is small, and essentially saturates for larger $N$.

\begin{figure}[!t]
    \centering
    \includegraphics[width=\columnwidth]{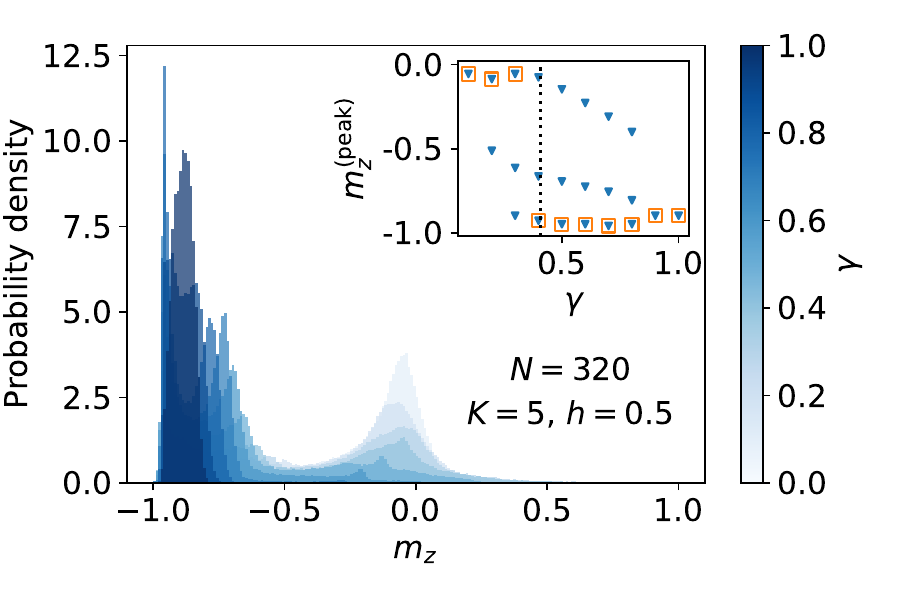}
    \caption{Same as in Fig.~\ref{fig:dist}, but for fixed $N=320$ and for many values of $\gamma$ (see the color bar on the right). The inset shows the position of the peaks of the probability distribution as a function of $\gamma$. The highlighted points correspond to the absolute maximum for that $\gamma$.
    }
    \label{fig:dist-manygamma}
\end{figure}

We finally study how the single peak bifurcates as the decay rate increases, for a fixed $K$. One example is shown in Fig.~\ref{fig:dist-manygamma}, where we plot the magnetization distributions obtained for $N = 320$ for $\gamma \in [0.1, 1.0]$, for $K = 5$. The inset reports the position of the peaks of the probability distribution as a function of the decay rate. The highlighted points surrounded by squares correspond to the highest peak for that $\gamma$. The vertical dashed line marks the transition between the semiclassical chaotic regime (left of the line) and the regular one (on the right). We see that the highest peak (global maximum) of the distributions shows a discontinuity at the chaotic-to-regular transition.

\section{Conclusions}\label{conc:sec}
We studied a dissipative quantum many-body system, whose thermodynamic limit corresponds to a classical mean-field dynamics and can show a chaotic behavior. Thanks to the fact that the finite-size system along quantum trajectories is in a pure many-body state, we evaluated measures of entanglement and quantum complexity, relating them with the properties of chaos/regularity (according to the sign of the largest Lyapunov exponent) in the thermodynamic limit.

Quantum magic is sensitive to the mean-field chaotic properties of the system. This quantity measures the complexity of the state---meant as distance from the set of stabilizer states---and its average over the trajectories reaches an asymptotic value in time, with clearly distinct behaviors in the classically chaotic and classically regular regimes.  In the chaotic case, the asymptotic nonstabilizerness shows constant plateaus in the system parameters, and without a coupling to the environment (unitary case) its behavior is compatible with a logarithmic scaling with the system size (i.\,e., with the dimension of the accessible Hilbert space), in agreement with the theoretical predictions~\cite{Leone2021quantumchaosis}. Switching on the coupling with the environment, there are still plateaus with the system parameters, but the behavior is consistent with a subextensive power-law scaling with the system size. In the regular regime, there is still a power-law behavior, but with no plateaus. Differently from the chaotic regime, the scaling exponent nontrivially depends on the system parameters (there are also intervals where the scaling exponent is near to one and the nonstabilizerness is extensive).

We have also considered the half-system entanglement entropy averaged over the trajectories. As the nonstabilizerness, this quantity reaches an asymptotic value, but its behavior has no apparent relation with classical chaos/ergodicity properties.
In particular, we observed a peculiar non-monotonic dependence with the system size, with a maximum reached at a critical size that depends on the system parameters and shows no discontinuities at the transitions from regular to chaotic behaviors. 
Finally, we studied the expectation of the magnetization along $z$: its variance along and across the trajectories reflects the underlying classical behavior, in the scaling with the system size.
In fact, in the chaotic case, the variance increases with the system size, such that the corresponding curves show an upper concavity with $N$.
Conversely, in the regular case, after an initial transient the variance settles down to a finite value, displaying a lower concavity with $N$.
This change of behavior reflects into a discontinuity in the global maximum of the full distribution of the expectations.

What is most remarkable in this analysis is that the thermodynamic-limit chaos is witnessed at finite size by a complexity measure (the magic) and not by an entanglement measure (the entanglement entropy). To understand whether this is an instance of a more general fact or not, future research may focus on the behavior of such quantities in strongly chaotic systems evolving along stochastic quantum trajectories.

\begin{acknowledgements}
We would like to thank R.\,Carter, R.\,Fazio, A.\,Hamma, and E.\,Tirrito for very helpful discussions and comments. G.\,P. and A.\,R. acknowledge financial support from PNRR MUR Project PE0000023-NQSTI. G.\,P. acknowledges computational resources from the CINECA award under the ISCRA initiative, and from MUR, PON “Ricerca e Innovazione 2014-2020”, under Grant No.~PIR01\_00011 - (I.Bi.S.Co.). This work was supported by PNRR MUR project~PE0000023 - NQSTI, by the European Union’s Horizon 2020 research and innovation programme under Grant Agreement No~101017733, by the MUR project~CN\_00000013-ICSC (P.\,L.),  and by the  QuantERA II Programme STAQS project that has received funding from the European Union’s Horizon 2020 research and innovation programme under Grant Agreement No~101017733 (P.\,L.). 
\end{acknowledgements}

\appendix

\section{Other probes of classical chaos}\label{other:sec}
\subsection{Bifurcation diagrams}
\begin{figure*}[!t]
    \centering
    \includegraphics[width = \textwidth]{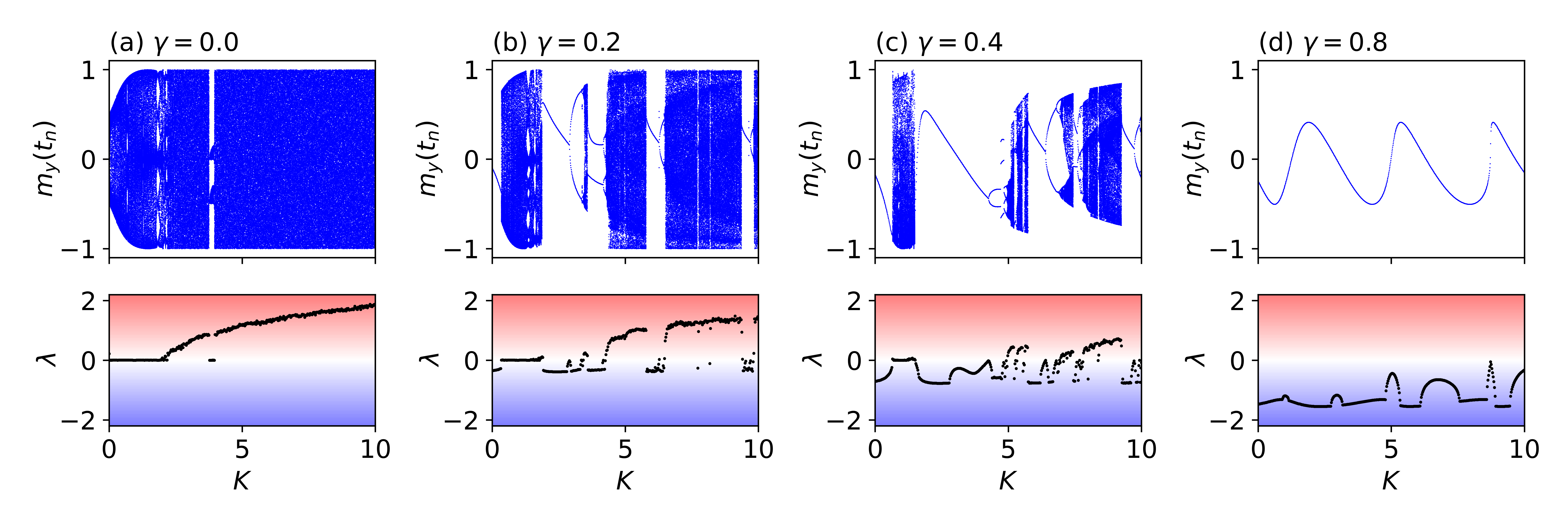}
    \caption{(top panels) Bifurcation diagrams: the last $M = 250$ stroboscopic values of $m_y$ as a function $K$, at the end of dynamics lasting $N_\tau = 10^4$ periods, for $h=0.5$ and four different values of $\gamma$ (see the labels). The initial condition is $\vec {m}(0) = (0, 0, 1)$. (bottom panels) The largest Lyapunov exponent as a function of the kick strength.  The background color follows the same color scheme as in Fig.~\ref{fig:lyapunov-exponent}.}
    \label{fig:mf-bifurcation}
\end{figure*}
Another way of studying the transition between chaotic and regular behavior in the mean-field classical limit is by looking at bifurcation diagrams, showing the stroboscopic values of the magnetization components as a function of $K$. In particular, when the dynamics converges to stationary points, the stroboscopic values taken after a transient are all equal to each other. When there is period doubling, the stroboscopic values split, producing bifurcations. Through a bifurcation cascade, the system may end up in a chaotic regime, where stroboscopic values cover dense ranges of possible values for $m_\alpha$; this is a universal route to chaos~\cite{Feigenbaum}. 

We can observe the bifurcation cascade in Fig.~\ref{fig:mf-bifurcation}, where we plot the last $M = 250$ stroboscopic values of $m_y$ at the end of the dynamics lasting $N_\tau = 10^4$ periods, as a function of $K$ and for different $\gamma$. In the bottom row, we report the corresponding largest Lyapunov exponent $\lambda$. We see that the succession of regular and chaotic behaviors can only be seen at relatively small values of $\gamma$, see panels (b-c): when the decay rate increases too much, the first bifurcation cascade leading to chaos moves to higher and higher values of $K$, until eventually it is no longer visible and the dynamics is always regular and converging to a stable fixed point, as confirmed by the fact that $\lambda<0$ [see panel (d)]. Moreover, when $\lambda=0$, the dense region observed in the bifurcation plot corresponds to a regular dynamics along the closed orbits displayed in Fig.~\ref{fig:mf-poincare}.

\subsection{Hausdorff dimension}

Consider the last column of Fig.~\ref{fig:mf-poincare}, referring to the kick strength $K = 5$. Increasing the decay rate, the system goes from a situation where the dynamics is fully ergodic and classical trajectories uniformly cover the Poincaré section, to an opposite limit where classical trajectories sample a very narrow region of the phase space, while they converge to the unique steady state of the dynamical system. In a sense, the effective dimension of the phase space section covered by the classical trajectories is approximately two (a square) in the first case and would be approximately one if all the points were on a line. In between, we observe a complicated scenario in which the points of the classical trajectories organize into fractal structures. The Hausdorff dimension $d_H$~\cite{Schuster,Ott} characterizes the effective dimension of these fractal structures. 

\begin{figure}[!t]
    \centering
    \includegraphics[width = \columnwidth]{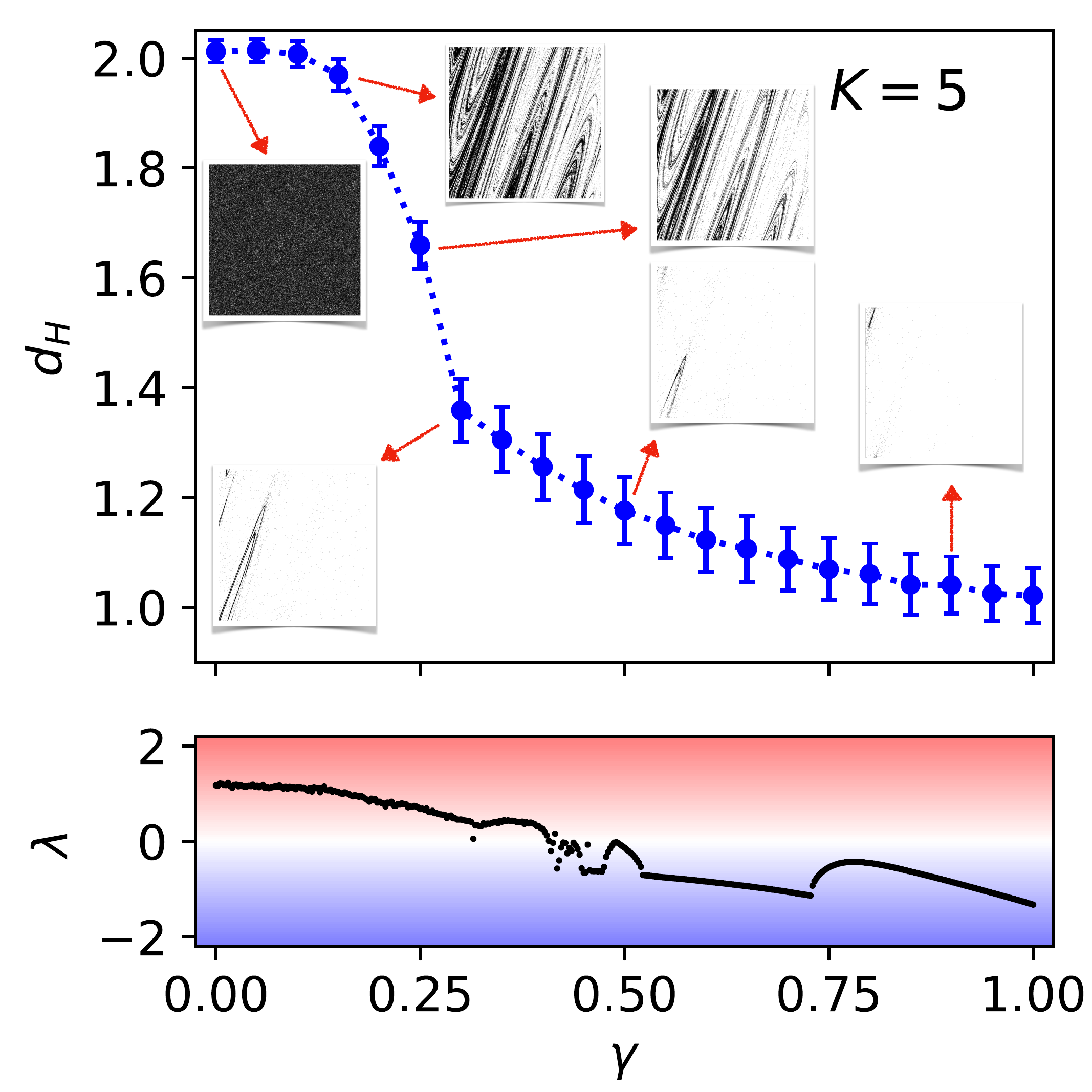}
    \caption{(Top) Estimated Hausdorff dimension $d_H$ of the mean-field dynamics versus $\gamma$. (Bottom) Largest Lyapunov exponent $\lambda$ versus $\gamma$. The background color follows the same color scheme as in Fig.~\ref{fig:lyapunov-exponent}. Parameters: $K = 5$, $h = 0.5$.}
    \label{fig:hausdorff}
\end{figure}

In order to estimate $d_H$, we divide the Poincaré section into a grid of non-overlapping squares with edges of length $\epsilon$. We then count how many squares $\mathcal{N}(\epsilon)$ we need to fully cover the blue points in each panel of Fig.~\ref{fig:mf-poincare}. Notice that, if the points were uniformly distributed in the entire phase space, we would need $ \mathcal{N}(\epsilon) \sim 1/\epsilon^2$ squares to cover them; instead, if they were on a line, $ \mathcal{N}(\epsilon) \sim 1/\epsilon$; therefore, in general, $ \mathcal{N}(\epsilon) \sim 1/\epsilon^{d_H}$.
The Hausdorff dimension $d_H$ is thus defined as
\begin{equation}\label{eq:hausdorff}
    d_H \equiv -\lim_{\epsilon\to0} \frac{\ln \mathcal{N}(\epsilon)}{\ln \epsilon}.
\end{equation}
In order to estimate it numerically, we consider $M_0 = 500$ initial conditions and consider 10 square sizes $\epsilon \in [2^{-10}, 2^{-1}]$.

In Fig.~\ref{fig:hausdorff}, we plot it versus $\gamma$, for $K = 5$. We can see that the Hausdorff dimension decreases by increasing $\gamma$ and saturates to a plateau corresponding to $d_H \approx 1$. This plateau starts to develop at $\gamma \approx 0.3$; the saturation is numerically complete (within error bars) at $\gamma \approx 0.5$. We also show the Poincaré sections corresponding to some of the considered values of $\gamma$, to make the trend easier to follow. Notice the drastic change between the Poincaré section at $\gamma = 0.25$ and the one at $\gamma = 0.3$, signaled by a large change in $d_H$. 

In the bottom part of the figure, we plot the Lyapunov exponent $\lambda$ as a function of $\gamma$. The $d_H \approx 1 $ plateau corresponds to a negative value of $\lambda$, while values of $d_H$ close to $2$ are associated with a positive $\lambda$. The region around the chaotic-to-regular transition point is not exactly captured by the Hausdorff dimension, due to numerical instabilities in its estimation through the box-counting algorithm.

\begin{figure}[!t]
    \centering
    \includegraphics[width = \columnwidth]{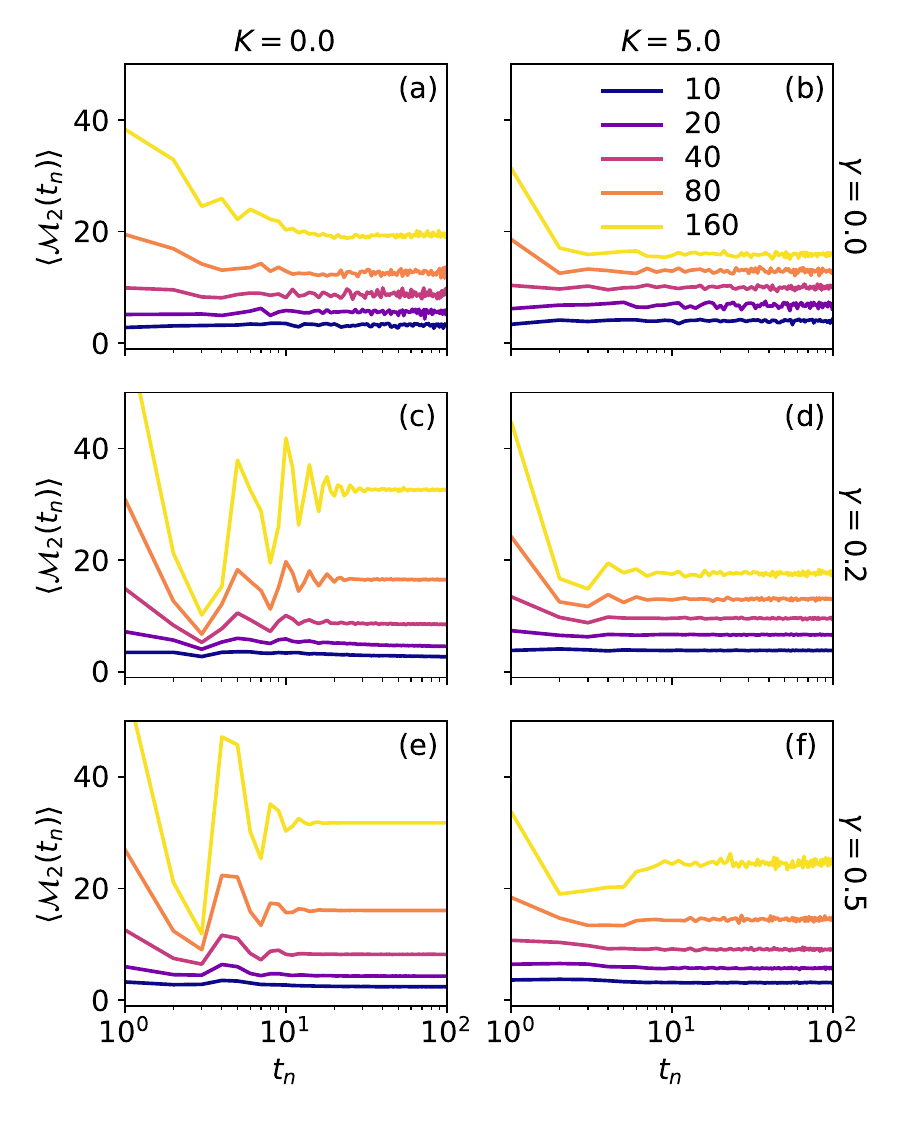}
    \caption{The stabilizer R\'enyi entropy $\langle \mathcal{M}_2(t_n) \rangle$, averaged over trajectories, versus the stroboscopic time $t_n$. The various colors stand for different sizes, from $N = 10$ (darkest) to $N=160$ (lightest), as indicated in the legend. Left and right columns are for $K=0$ and $K=5$, respectively. Rows from top to bottom: $\gamma=0, \, 0.2, \, 0.5$. We fix $h = 0.5$.}
    \label{fig:sre-dynamics}
\end{figure}

\section{\add{Additional numerical results}}\label{app:numerics}

In this section, we provide additional numerical results concerning nonstabilizerness and entanglement dynamics to accompany the results discussed in the main text.

\subsection{Nonstabilizerness}

Figure~\ref{fig:sre-dynamics} reports the behavior of the nonstabilizerness $\langle \mathcal{M}_2(t_n) \rangle$ versus the stroboscopic time $t_n$ in the dissipative quantum kicked top, after averaging over the randomness of the quantum trajectories.
We first notice that the SRE as a function of time rapidly tends to an asymptotic value. Moreover, when the kick strength $K$ is small and the dynamics is regular, the asymptotic value seems to grow extensively with $N$ (left panels, for $K=0$). On the other hand, for larger kick strengths, the dynamics becomes chaotic and the asymptotic value increases more slowly, possibly as $\ln N$ (right panels, for $K=5$).

\begin{figure}[t]
    \centering
    \includegraphics[width = \columnwidth]{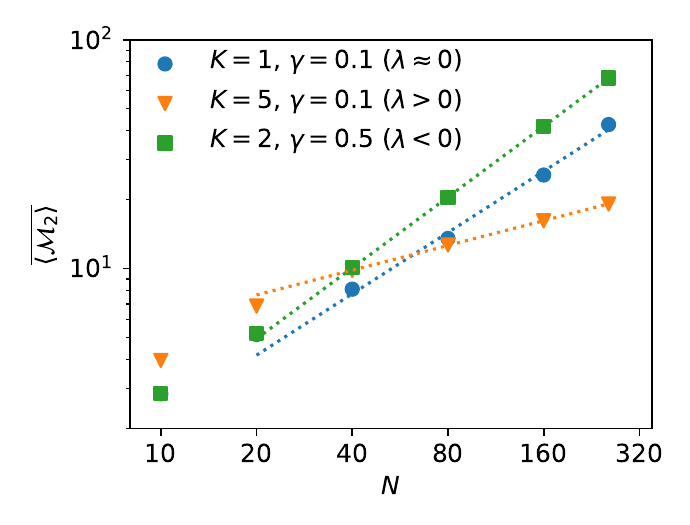}
    \caption{Asymptotic SRE versus $N$, for various parameters that corresponds to different regimes of the semiclassical dynamics. Dotted lines denote the fitting functions $\overline{\langle \mathcal{M}_2\rangle} \sim N^\alpha$. In our fit we exclude the smallest size $N = 10$. Notice the double-logarithmic scale. We fix $h=0.5$.}
    \label{fig:sre-N}
\end{figure}

To be more quantitative, we first extrapolate the asymptotic value of the SRE as discussed in Sec.~\ref{traj:sec}. We plot it versus the system size (up to $N = 256$) in Fig.~\ref{fig:sre-N}, for various combinations of the kick strength $K$ and the decay rate $\gamma$, corresponding to different values of the Lyapunov exponents of the undarlying semiclassical dynamics [see Fig.~\ref{fig:lyapunov-exponent}(a)]. We observe that, for parameters that correspond to regular semiclassical dynamics ($\lambda \le 0$), the asymptotic SRE is extensive and grows as $\overline{\langle \mathcal{M}_2\rangle} \sim N^\alpha $, with $\alpha $ close to one. On the other hand, in the chaotic regime, the SRE grows more slowly. Indeed, a numerical fitting returns an exponent that is appreciably smaller than one (see Tab.~\ref{tab:sre-scaling}). Given the relatively small available sizes, by simply analyzing these data, we cannot rule out the possibility that the scaling is sub-extensive (but still power law), rather than logarithmic, which is why, in the main text, we focus on the more interesting scaling of $\delta\mathcal{M}_N$, see Sec.~\ref{nosta:sec}.

\subsection{Entanglement entropy}

\begin{figure}[t]
    \centering
    \includegraphics[width = \columnwidth]{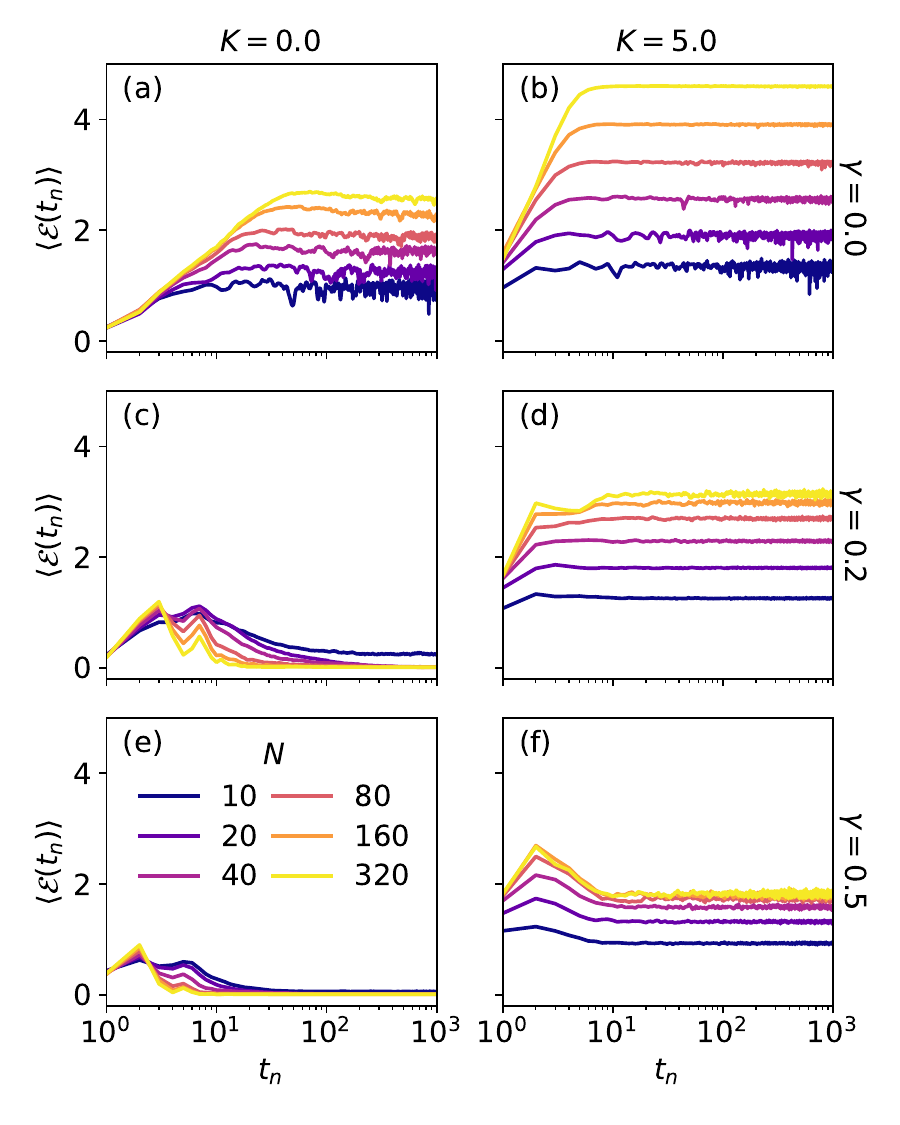}
    \caption{The half-system entanglement entropy $\braket{\mathcal{E}(t_n)}$, averaged over the trajectories, versus the stroboscopic time $t_n$. 
    The colored curves are for different sizes, from $N = 10$ (darkest) to $N=320$ (lightest). The values of $K$ and $\gamma$ are indicated in the labels of the panels (same pattern as in Fig.~\ref{fig:sre-dynamics}). We fix $h = 0.5$.}
    \label{fig:entropy-dynamics}
\end{figure}

In Fig.~\ref{fig:entropy-dynamics} we show some examples of the EE $\braket{\mathcal{E}(t_n)}$, averaged over the trajectories, versus the stroboscopic time $t_n$.
The top row corresponds to the unitary limit $\gamma = 0$, for the two values of the kick strength $K = 0$ where the mean-field dynamics is regular [panel (a)], and $K = 5$ where it is chaotic [(panel (b)]. All the entropies increase and then saturate to the asymptotic values.
To be more quantitative, it is useful to focus on the properties of the asymptotic value $\overline{\langle \mathcal{E} \rangle}$. In the ergodic regime, we find $\overline{\langle \mathcal{E} \rangle} = \alpha + \beta \ln (N + 1)$, with $\alpha = -1.05\pm 0.07$ and $\beta = 0.974\pm 0.016$ for $K=5$. The fact that $\beta \approx 1$ agrees with the expectation that the EE reaches its maximum value $\mathcal{E}_\text{max} = \ln D$ (where $D=N+1$),
for an ergodic dynamics. In contrast, when $K$ is smaller, we observe a slower growth with the system size: using the same fitting function, for $K=0$ we get $\alpha = -0.17\pm 0.07$ and $\beta = 0.477 \pm 0.016$.

\begin{figure}[t]
   \centering
    \includegraphics[width=\columnwidth]{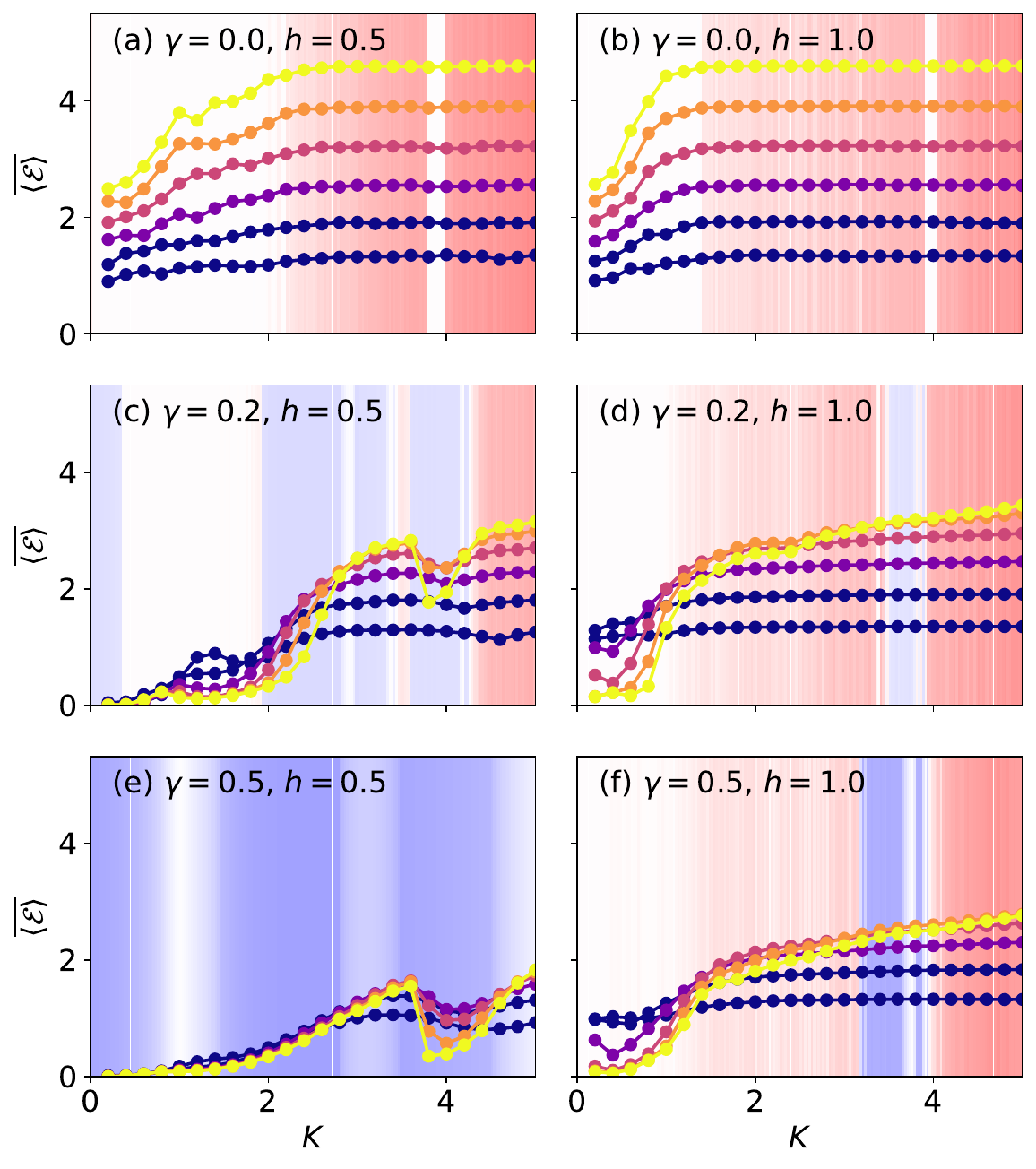}
    \caption{The asymptotic entanglement entropy $\overline{\langle \mathcal{E} \rangle}$ versus $K$, for different sizes $N \in \lbrace 10, 20, 40, 80, 160, 320 \rbrace$ (the legend is the same as in Fig.~\ref{fig:entropy-dynamics}), and for several combinations of $\gamma$ and $h$, as indicated in the labels of the panels. The background color is related to the value of the largest Lyapunov exponent $\lambda$, following the same color code as in Fig.~\ref{fig:lyapunov-exponent}.}
    \label{fig:entropy-kick-scaling}
\end{figure}

In the central and lower rows of Fig.~\ref{fig:entropy-dynamics}, we show the EE versus time in the presence of dissipation, being it either weak ($\gamma = 0.2$, middle row) or strong ($\gamma = 0.5$, bottom row), for the same values of $K$ considered before. Panels (c),(e), and (f) correspond to a regular mean-field dynamics ($\lambda<0$), while panel (d) is for a chaotic dynamics ($\lambda>0$). We observe that, after a transient, all curves converge to an asymptotic value.
Further insight can be obtained by focusing on the dependence of the asymptotic value $\overline{\langle \mathcal{E} \rangle}$ on $K$.

In Fig.~\ref{fig:entropy-kick-scaling} we show the behavior of $\overline{\langle \mathcal{E} \rangle}$ versus $K$, for the same values of $\gamma$ and system sizes used in Fig.~\ref{fig:entropy-dynamics}. First of all, we notice a weak relation between such curves and the sign of the largest Lyapunov exponent $\lambda$ (see the color code in the background, as in Fig.~\ref{fig:sre-diff-scaling}). Furthermore, for increasing $N$, the values of $\overline{\langle \mathcal{E} \rangle}$ display a tendency to saturate to a finite limit (area-law behavior) and even to decrease.

In the main text, we focus our attention on the scaling of the asymptotic value of the EE with $N$.

\bibliographystyle{quantum}

\end{document}